\documentclass[a4paper,11pt]{article}
\pdfoutput=1
\usepackage{jcappub}
\usepackage[T1]{fontenc}
\usepackage{amsmath}
\usepackage{amssymb}
\usepackage{makecell}
\usepackage{graphicx}
\usepackage{subfig}
\usepackage{subcaption}
\usepackage{array}
\usepackage{placeins}

\usepackage{fontawesome5}
\usepackage{tabularx}
\makeatletter
\newcommand{\github}[1]{%
    \href{#1}{\faGithub}%
}
\makeatother

\title{Segmenting proto-halos
with vision transformers}

\author{Toka Alokda,}
\author{Cristiano Porciani}
\affiliation{Argelander-Institute for Astronomy, University of Bonn,
Auf dem Hügel 71, 53121 Bonn, Germany}

\emailAdd{talokda@astro.uni-bonn.de}
\emailAdd{porciani@astro.uni-bonn.de}

\abstract{The formation of dark-matter halos from small cosmological perturbations generated in the early universe is a highly non-linear process typically modeled through N-body simulations. In this work, we explore the use of deep learning to segment and classify proto-halo regions in the initial density field according to their final halo mass at redshift $z=0$. We compare two architectures: a fully convolutional neural network (CNN) based on the V-Net design and a U-Net transformer. We find that the transformer-based network significantly outperforms the CNN across all metrics, achieving sub-percent error in the total segmented mass per halo class. 
Both networks deliver much higher accuracy than the perturbation-theory-based model \textsc{pinocchio}, especially at low halo masses and in the detailed reconstruction of proto-halo boundaries.
We also investigate the impact of different input features by training models on the density field, the tidal shear, and their combination. Finally, we use Grad-CAM to generate class-activation heatmaps for the CNN, providing preliminary yet suggestive insights into how the network exploits the input fields.}

\arxivnumber{2508.00049}

\begin{document}

\maketitle

\section{Introduction}

It is widely accepted that
the large-scale structure (LSS) of the Universe arises from the non-linear gravitational collapse of small primordial density perturbations seeded during the inflationary epoch.
This process results in the formation of a population of virialized dark-matter (DM) halos, which provide the environments where galaxies form \citep[e.g.][]{WhiteandRees_1978}.

A fundamental question in this context is: which regions in the initial conditions (Lagrangian space) evolve into DM halos --- or even galaxies --- with a specific set of properties at a given cosmic time?
Following \cite{PDH-2002a}, we refer to such regions as proto-halo patches or simply proto-halos. In N-body simulations,
they can be readily identified
by tracking the computational elements (particles)  associated with a final bound structure back to their initial positions \cite{PDH-2002b}. This backward mapping
reveals that proto-halos are typically elongated, approximately ellipsoidal regions characterized by converging mass flows. Their shape tends to align with the direction of maximum compression in the local tidal field \cite{lee-pen, PDH-2002b, ludlow-porciani-2011, Despali+2013, Ludlow+14, Miko+14}.
Notably, proto-halos differ significantly from regions enclosed by
iso-density surfaces in the initial field. Their boundaries
are fuzzy rather than sharply defined,
a feature that is partly shaped by the particular halo definition used in the analysis (see e.g. figures 4 and 11 in \cite{Ludlow+14}).

A natural next step is the development of a predictive model capable of accurately identifying proto-halos without the need for full N-body simulations.
 If such a model were available, it would enable a wide range of applications. For instance, it could be used to efficiently select regions for high-resolution zoom-in simulations \cite[e.g.][]{KW-1993, Bert-2001, Hahn_2011}
or to inform priors on large-scale (Lagrangian) bias parameters in perturbative analyses of halo clustering \cite[e.g.][]{Kaiser-1984, CPK-2000, Matsubara-2008, Vlah+16, Modi+2017, Zennaro+2022}. 
When combined with a low-resolution simulation, such a model would also offer a computationally efficient method to generate large ensembles of mock halo or galaxy catalogs \cite[e.g.][]{PP-1996, monaco-2016}.

From a theoretical standpoint, 
identifying proto-halos is a problem of intrinsic interest,
as it encapsulates the complexities 
of gravitational collapse and
its environmental dependence within the cosmic web.
Traditionally, this challenge has been addressed through a two-step approach.

To begin with, one adopts a model for the monolithic collapse of an isolated perturbation, neglecting the influence of internal substructures.
Several collapse models have been proposed for this purpose. The most widely used is the spherical collapse model, which assumes that the evolution of a perturbation depends solely on the mean overdensity within a spherical region \citep{GG-72}. A more refined description is provided by the ellipsoidal collapse model of \cite{PP-1996}, in which an initially spherical patch becomes anisotropic under the influence of external tidal fields. An alternative ellipsoidal model, proposed by \cite{Ludlow+14}, starts from initially triaxial perturbations, capturing additional geometric complexity.

The second step consists of applying an algorithm to identify regions in Lagrangian space where the chosen collapse model can be meaningfully applied. Over the past few decades, two primary frameworks have emerged for this task.
In the so-called peak theory \cite[e.g.][]{BBKS-1986}, DM halos are assumed to form around local maxima of the linearly extrapolated overdensity field, smoothed on a mass scale corresponding to the desired halo mass. The collapse model is then applied to the region
surrounding the identified peak, which can be isolated using various criteria.
While simulations confirm that this halo–peak association holds for the majority of halos \citep{ludlow-porciani-2011, Hahn-Paranjape-2014}, a substantial population of low-mass halos is found to accumulate significantly more mass than peak theory would predict \citep{ludlow-porciani-2011}.

An alternative and widely used approach, originally introduced by Press \& Schechter \citep{pressshechter}, posits that a given mass element collapses into a halo if the linearly evolved, smoothed overdensity at its Lagrangian position exceeds a collapse threshold set by the chosen collapse model. In the standard cold dark matter (CDM) scenario, where perturbations exist on all scales, this idea requires refinement to account for the interplay of fluctuations across different smoothing radii. This is addressed by the excursion-set formalism \citep[e.g.][]{bondetalPS, zentner-2007}, which follows the “trajectory” of the overdensity $\delta(R)$
at a fixed Lagrangian position as a function of the smoothing scale $R$.
In this framework, a halo is identified with the first up-crossing of a collapse threshold, which can be modeled as environment-dependent \citep{SMT-2001, ZOMG_I} or include a stochastic component to capture effects beyond the simplified collapse model \citep{MR-2010}. When applied to the full linear density field, the excursion-set approach can be fine-tuned to match the halo mass function observed in simulations \citep[e.g.][]{SMT-2001, Rob+2009, Elia+2012, Ludlow+14}. However, detailed comparisons show that it yields poor mass estimates for the halo in which a randomly chosen particle ends up \citep{White-1994}. In contrast, predictions improve significantly when the method is applied to particles near the proto-halo’s center of mass \citep{SMT-2001}.
This limitation has motivated the development of hybrid models that combine elements of the excursion-set framework with peak theory, aiming to leverage the strengths of both approaches \citep[e.g.][]{PeaksES-2012, SSM-2021}. An alternative approach proposes that halos form around peaks of the smoothed potential energy, rather than in the density field itself \cite{Musso-Sheth-2021}.

By adapting and extending the theoretical frameworks discussed above,
several numerical algorithms have been developed to 
generate approximate catalogs of DM halos from realizations of the linear density field, often at negligible computational cost.
These  methods generally proceed in two main stages: 
(i) identification of proto-halos in the initial conditions;
(ii) displacement of these patches to their final
(Eulerian) positions using Lagrangian perturbation theory (LPT).
For the identification step,
the peak-patch method \cite{PP-1996,mpp-2019} searches for the 
largest spherical regions in Lagrangian space that satisfy
the conditions for ellipsoidal collapse at a given redshift.
This procedure includes rules for hierarchical exclusion and merging to prevent overlapping halo assignments.
In contrast, the code 
\textsc{pinocchio} \cite{pinocchio,Monaco_2002} 
uses perturbative techniques to 
estimate the time at which individual particles first undergo orbit crossing, marking the onset of collapse. 
These particles are then grouped into halos based on their collapse times and spatial proximity, with the grouping procedure calibrated to reproduce the halo population observed in full N-body simulations.

With the rise of artificial intelligence (AI), new methodologies are emerging that leverage N-body simulations as training data to bypass explicit physical modeling. One notable example is the generalization of the excursion-set formalism into a machine learning framework that does not require a predefined collapse model \citep{luciesmith2018}.
In this approach, a random forest (RF) binary classifier \cite{randomforests}
is trained to predict whether a given N-body particle --- i.e. the matter contained in a voxel in the initial conditions --- will end up in a halo above a specified mass threshold at redshift $z=0$. 
The input features are derived from the density trajectory
$\delta(R)$, evaluated at the Lagrangian position of the particle. 
The model performs comparably to standard collapse-based approaches, and interestingly, adding tidal field descriptors such as ellipticity and prolateness does not yield a significant gain in accuracy \citep{luciesmith2018}.
The RF classifier has also been shown to outperform other binary classification algorithms \citep{chacon1} and remains effective even in modified gravity scenarios with a fifth force \citep{mlmodifiedgravity}. Furthermore, this methodology can be extended into a regression task using gradient-boosted trees
\citep{GBT1, GBT2, GBT3} to predict the final mass of the halo associated with each particle at $z=0$ \cite{luciesmith2019interpretable}. 

While the examples discussed above provide valuable starting points that highlight the potential of machine learning (ML) in this domain, they offer limited insight into the physical processes driving structure formation.
By design, the input features used in these models are carefully selected to emulate the extended Press–Schechter framework.\footnote{Originally introduced by \cite{bondetalPS} to derive the halo mass function from statistical principles, the extended Press–Schechter approach performs poorly at predicting the fate of individual particles. This is especially true for those located near the boundaries of proto-halos, where spherical smoothing kernels fail to properly capture the structure of collapsing perturbations \citep{White-1994, SMT-2001}.} Moreover, these features are typically reduced to flattened one-dimensional trajectories, which discard the rich three-dimensional spatial information contained in the initial fields.

This limitation can be addressed through deep learning, and in particular by using convolutional neural networks (CNNs) \citep{CNN}, a class of models that has revolutionized computer vision and image processing. CNNs learn to extract meaningful patterns from structured data by convolving the input with learned filters and applying non-linear transformations to generate increasingly abstract representations. Recent studies have employed CNNs to investigate various aspects of halo formation \citep{vnet_berger, vnet_bernardini, luciesmith2024deep, vnet2024segment,buisman2025differentiablehalomassprediction}.

In the context of CNNs, identifying the matter that will collapse into halos corresponds to partitioning 3D Lagrangian space into semantically meaningful regions --- a task analogous to semantic segmentation in computer vision. The simplest formulation casts this as a binary classification problem: distinguishing between mass elements that will end up in halos and those that will not. A CNN can then output a confidence score for each voxel in the Lagrangian grid, which quantifies the network’s certainty regarding the classification \citep{vnet_berger}. Alternatively, the classification can be reformulated as a distance-based regression task
by applying the Euclidean Distance Transform to convert binary region labels into continuous distance fields \citep{vnet_bernardini}.

Recent work has shown that the V-Net architecture, originally developed for 3D medical image segmentation \citep{vnet}, performs remarkably well in predicting proto-halo regions. Specifically, V-Net has been successfully applied to reproduce proto-halos identified by the peak-patch model \citep{vnet_berger} and by full N-body simulations \citep{vnet_bernardini, vnet2024segment,buisman2025differentiablehalomassprediction}. Alternatively, CNNs can be trained to predict the final
mass of the halos that host randomly selected N-body particles \cite{luciesmith2024deep}.
Percent-level accuracy can be achieved when the analysis is limited to Lagrangian voxels located at the actual centers of proto-halos, particularly in the case of massive halos \citep{LS-B-S-2023}.

The ML-based models discussed above operate at the level of individual particles or voxels and are not designed to identify proto-halo patches as distinct, coherent units.

To construct mock halo catalogs, however, it is essential to go beyond voxel-wise classification and partition the Lagrangian volume --- previously labeled as containing proto-halo material --- into separate, well-defined objects. This fragmentation step is critical for associating mass elements with specific halos and for generating catalogs that reflect realistic halo populations. A variety of strategies have been proposed to address this challenge. These include custom methods for detecting connected components in the predicted proto-halo regions \citep{vnet_berger}, adaptations of the watershed algorithm to the 3D Lagrangian setting \citep{vnet_bernardini}, and more recent applications of instance segmentation techniques \citep{vnet2024segment}.

While these approaches have produced encouraging results, there remains substantial room for improvement, particularly in accurately separating neighboring structures and preventing spurious fragmentation or merging of proto-halos. These issues are clearly illustrated in, for example, figure 7 of \cite{vnet2024segment}.

Transformer models \citep{attentionisallyouneed} have initiated a paradigm shift in AI. Originally developed for natural language processing (NLP), these neural network architectures rely on mechanisms known as attention and self-attention to capture contextual relationships within sequential data --- such as the dependencies among words in a sentence. 
 When adapted to computer vision tasks, these models are referred to as vision transformers (ViTs). In this framework, input images are divided into smaller, fixed-size patches that are treated analogously to tokens in NLP. By leveraging self-attention, ViTs can model long-range interactions and capture global context across the input, often outperforming CNNs in tasks where such holistic understanding is crucial.
 
Over the past few years, ViTs have emerged as powerful alternatives to CNNs in image analysis, particularly when trained on large datasets. However, despite their growing success, their application to proto-halo detection remains unexplored. 

In this work, we aim to bridge that gap. As a first step, we frame proto-halo prediction as a classification problem into several mass bins (plus a non-halo class), rather than a full regression. This choice serves as a proof of concept: it reduces computational demands and dataset requirements, while providing a clear benchmark for testing deep learning architectures on this task. We train two types of networks to implement this idea: a CNN based on the V-Net architecture, and a U-Net Transformer (UNETR) \citep{unetr}. Both take as input the initial density field at redshift $z=99$ and output a multi-channel 3D probability mask, where each voxel is assigned a value between 0 and 1, indicating the model’s confidence that it belongs to a proto-halo of a given mass range. Demonstrating success in this simpler framework lays the foundation for future work extending to regression, which will require larger training sets and more powerful hardware.

The essential structure of the paper is as follows: 
methods are detailed in section~\ref{sec:methods}, results are presented in section~\ref{sec:results}, and conclusions are provided in section~\ref{sec:summary}.

\section{Methods}\label{sec:methods}
\subsection{N-body simulations and proto-halo definition}

We use N-body simulations performed with \textsc{Gadget-2} \cite{gadget}
to study structure formation
in a flat $\Lambda$CDM scenario which is consistent
with the final results of the Planck mission \cite{planck2018}.
The cosmological background is characterized by the density
parameters $\Omega_\mathrm{m}=0.315$, $\Omega_\Lambda=0.685$, and $\Omega_\mathrm{b}=0.049$,
while the present-day value of the Hubble constant is $H_0=67.4$ km $\mathrm{s^{-1}}$ Mpc. 
Scalar perturbations have a primordial spectral index of 
$n_s=0.965$ and are normalised such that $\sigma_8=0.811$.
We run four simulations within a periodic cubic box of sidelength $L=100 \,h^{-1}\, \mathrm{ Mpc}$, each containing 
$N=512^3$ particles of mass $6.5\times 10^8\,h^{-1}$ M$_\odot$.
The initial conditions at $z=99$ are generated using the \textsc{MUSIC} code \cite{Hahn_2011}.

We identify DM halos at $z=0$  
using the \textsc{Amiga Halo Finder} (AHF) \cite{AHF},
which selects spherical regions with a mean overdensity of 200 times the background density. We only consider halos containing 40 particles or more. 
 We obtain the proto-halos by tracing the halo particles back to their Lagrangian position at $z\rightarrow\infty$.

\begin{figure}
\centering
    \begin{tabular}{c c}
        \includegraphics[width=0.4\textwidth]{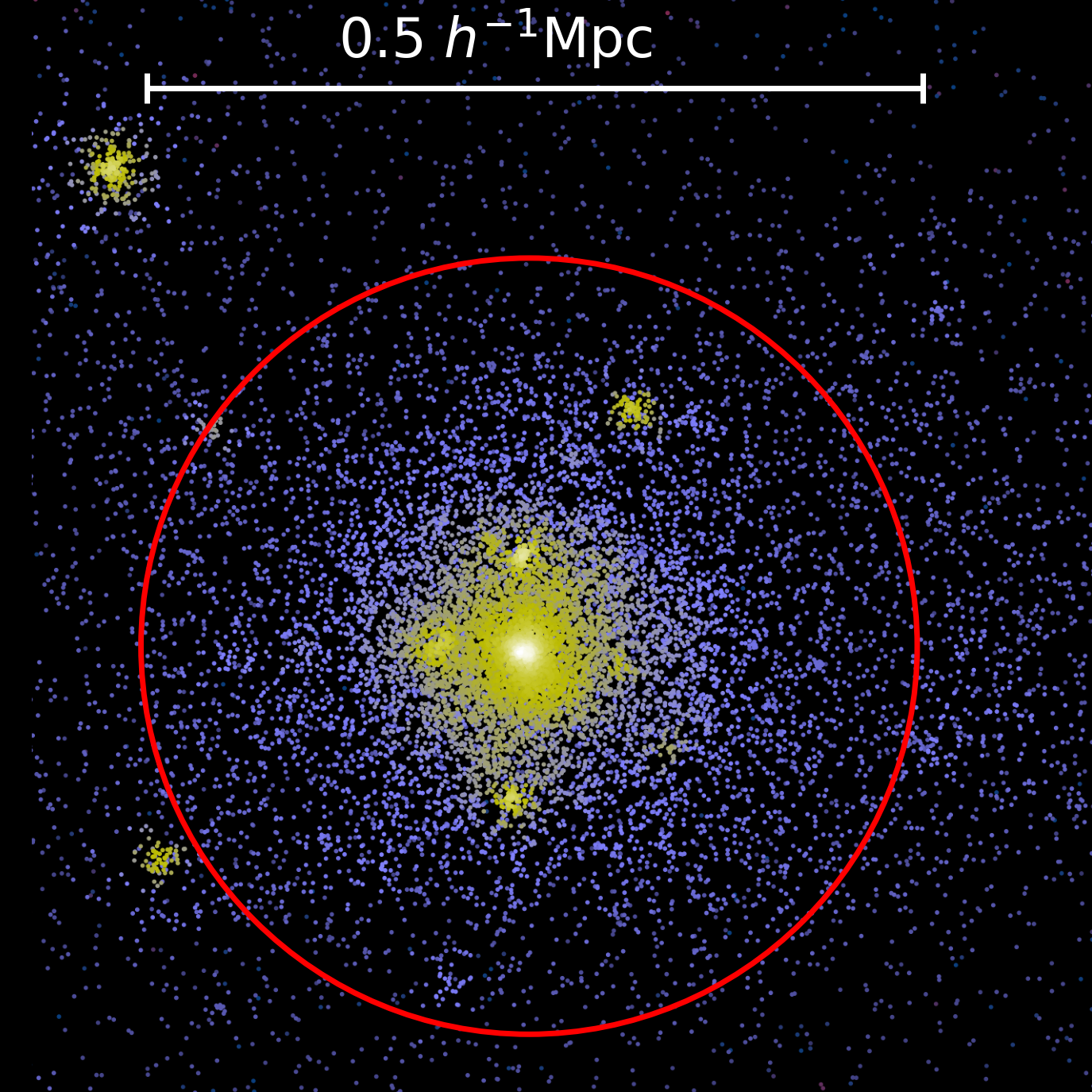} &
        \includegraphics[width=0.402\textwidth]{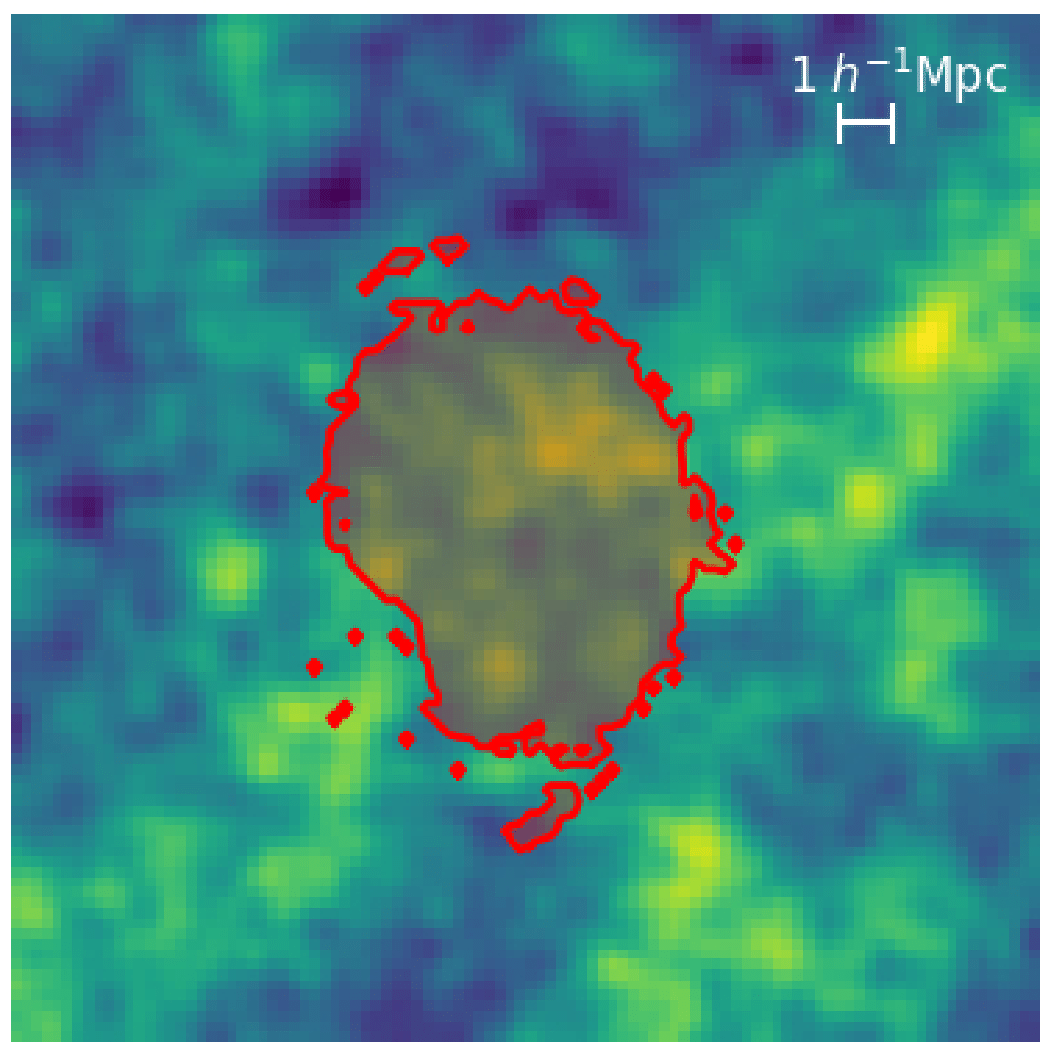}
    \end{tabular}
    \caption{Left: A dark-matter halo of mass $M=6.46\times 10^{12}\,h^{-1}$ M$_\odot$ at $z=0$,
    shown along with its immediate surroundings projected over a depth of $0.58\,h^{-1}\,\text{Mpc}$, corresponding approximately to the size of the structure. The red circle marks
    the outer boundary of the halo, defined as the radius enclosing  
    a mean over-density of $\Delta=200$. 
    The color of the N-body particles
    represents the local mass density, increasing logarithmically from blue to yellow, as estimated
    using an adaptive Gaussian kernel that contains 32 neighbors within its core.
    Right: The corresponding proto-halo patch at $z=99$
    (shaded region with thick red boundary), overlaid
    on the projected linear over-density field with a depth of $9.37\,h^{-1}\,\text{Mpc}$, chosen to encompass the full Lagrangian extent of the proto-halo.
    Bright yellow and dark blue indicate the highest and lowest density regions, respectively. 
    Note the difference in scales between the two panels: the material that initially spans several $h^{-1}$ comoving Mpc collapses into a compact structure of approximately $0.5\,h^{-1}$ Mpc in diameter.}
    \label{fig:halos}
\end{figure}

Figure~\ref{fig:halos} illustrates an example of a halo and its corresponding proto-halo.
For later analysis,
each proto-halo particle is tagged with the mass of its associated halo. 

\subsection{Data preparation}
\label{sec:dataprep}
We use the simulations and halo catalogs described above to construct three distinct datasets, each defined by a different binning scheme for the final halo masses (see table~\ref{tab:massbins} for details). 
Because the number density of halos decreases steeply with increasing mass, 
the highest mass bin is kept fixed across all datasets, as it contains only 150 to 162 halos per simulation box and thus provides limited statistical support for further subdivision.
The remaining bins are refined progressively from dataset 1 to dataset 3, with mass intervals becoming narrower in order to provide a more detailed resolution of the halo mass distribution.\footnote{In dataset 3, the average number of halos per simulation box decreases from 75{,}139 in the lowest mass bin to 156 in the highest. The intermediate bins contain 26{,}424, 9{,}909, and 939 halos, respectively.}
This incremental refinement allows us to assess how the classification performance of our models responds to increasingly fine-grained mass differentiation.

The resulting datasets exhibit a pronounced class imbalance: on average, the non-halo class alone accounts for 64.1\% of all voxels, while the halo classes combined make up the remaining 35.9\%. 
Moreover, the voxel distribution across halo mass bins is uneven. 
For instance, the class corresponding to the most massive halos contains 9.2\% of the voxels, while the lightest halo class in dataset 3 contains only 3.7\%.
This trend arises because massive halos, though rare, occupy disproportionately large Lagrangian volumes. Such voxel-wise imbalances are typical of segmentation problems and pose significant challenges during training, often requiring strategies such as data augmentation or loss weighting to mitigate bias toward dominant classes.

We subdivide each simulation box into 3375 overlapping sub-volumes of size $64^3$ voxels, corresponding to a comoving scale of $l = 12.5\, h^{-1} \mathrm{Mpc}$. This choice strikes a balance between resolution and efficiency: larger sub-volumes (e.g., $128^3$ voxels) would provide more structural context but demand substantially more memory and severely limit the feasible batch size, while significantly smaller sub-volumes would lose critical information, since $l$ is already comparable to the Lagrangian extent of the most massive proto-halos in our dataset. Although smaller proto-halos occupy only a limited fraction of voxels, they are far more numerous and thus appear frequently in the training set. In contrast, the rarer, larger proto-halos are underrepresented. The use of overlapping sub-volumes helps mitigate this imbalance by allowing the same massive proto-halo to be sampled in multiple sub-volumes, effectively boosting its statistical weight in training. This overlap acts as a form of data augmentation, exposing the model to varied spatial contexts of rare proto-halos and improving its ability to learn robustly across the full mass spectrum.

\begin{table}
    \centering
        \begin{tabular}{c c c c c c c}
        \hline
             Dataset & \multicolumn{6}{c}{Halo mass bin}\\
             & 1 & 2 & 3 & 4 & 5 & 6 \\
             \hline
             1 & 10.4 -- 11.4 & 11.4 -- 12.4 & 12.4 -- 13.4 & >13.4 & - & - \\
             2 & 10.4 -- 11.2 & 11.2 -- 12.0 & 12.0 -- 12.8 & 12.8 -- 13.4 & >13.4  & - \\
             3 & 10.4 -- 11.0 & 11.0 -- 11.6 & 11.6 -- 12.2 & 12.2 -- 12.8 & 12.8 -- 13.4 & >13.4\\
             \hline
        \end{tabular}
            
        \caption{Halo mass binning schemes used to construct the three datasets. Bin edges are given  in units of $\log_{10}\left( \frac{M}{h^{-1} \mathrm{M_\odot}} \right)$.}          
        \label{tab:massbins}
    
\end{table}

\subsection{Neural network architectures}
We employ two deep neural networks to predict proto-halo regions directly
from the initial conditions of N-body simulations.
Both models operate in a supervised learning framework, meaning that they are trained on paired input-output data,
where the input is the initial density field and the output is a labeled mask identifying proto-halo particles.
Once trained, the models are evaluated on independent validation and test datasets to assess their generalization performance.

The networks are designed to assign probability scores to individual N-body particles, each represented as a voxel in the Lagrangian space at $z=\infty$.  Based on these scores, predicted proto-halo regions are categorized into 
$P-1$ discrete mass bins, with an additional class for particles that are not expected to end up in any halo by redshift $z=0$.
The output is therefore four-dimensional: a three-dimensional probability mask with $P$ channels, corresponding to the $P-1$ mass bins and the non-halo class. The latter
also corresponds to the complement of a binary mask that combines all the mass-bin classes.

Both models belong to the class of volumetric deep learning networks, which process 3D volumes as both input and output. 
In the following subsections, we describe
their architectures and outline the training, validation, and testing procedures.

\subsubsection{\textsc{copra}: a fully-convolutional neural network}
\begin{figure}
    \centering
    \includegraphics[scale=0.055]{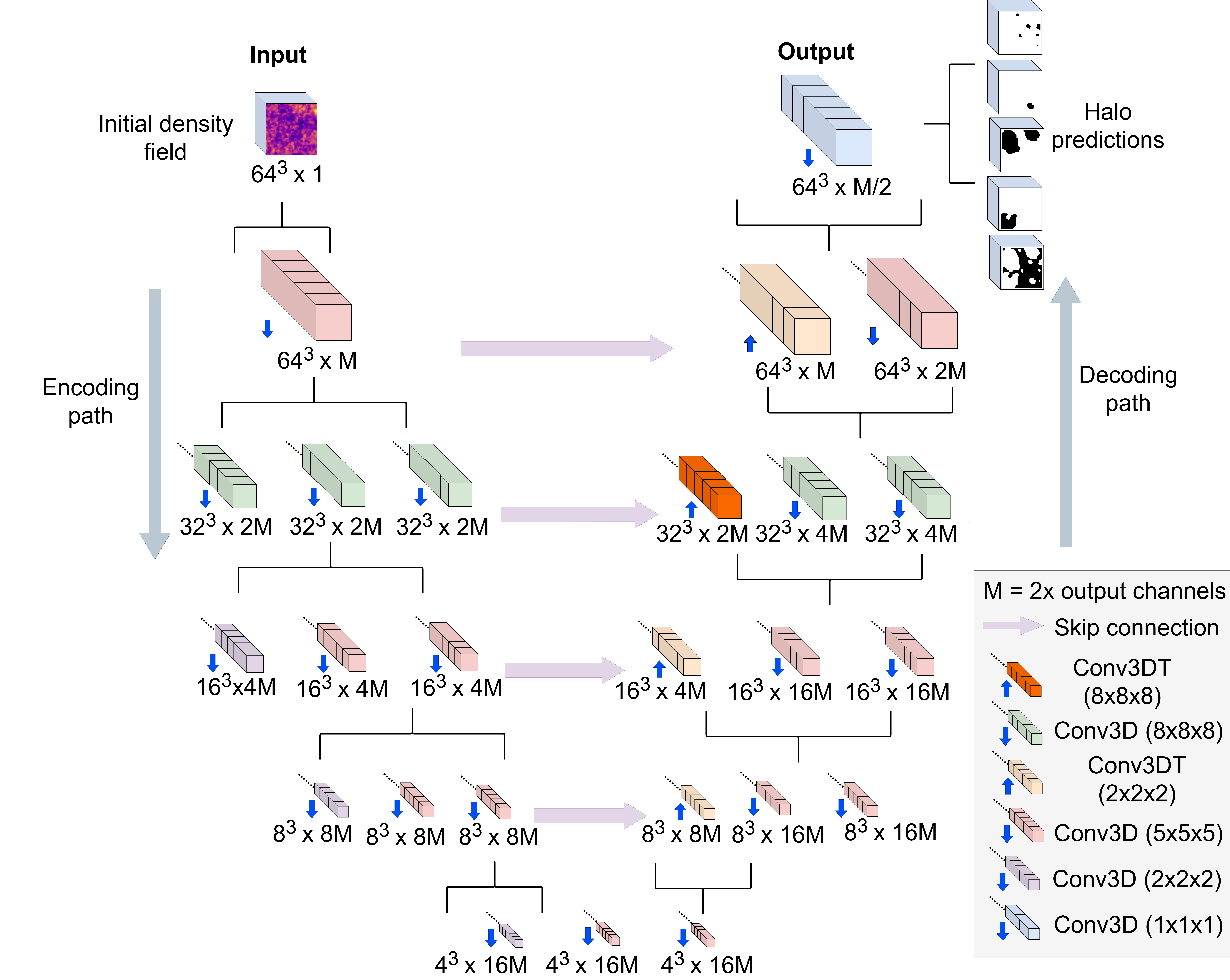}
    \caption{Architecture diagram of the V-Net model \textsc{copra} used in this study. 
    The network is illustrated for the case of 
$P=5$ output classes, corresponding to $M=2P=10$ base filters. The encoder–decoder structure includes convolutional and transposed-convolutional layers arranged in multi-layer blocks with skip connections, enabling effective learning of both global context and fine-grained spatial detail.}
    \label{fig:vnetfigure}
\end{figure}

Our first model --- which we dub \textsc{copra}, short for ``COnvolutions for PRoto-hAlo identification'' --- is a fully-convolutional neural network based on the V-Net 
architecture, an extension of U-Net to three-dimensional data. Prior studies have demonstrated that this class of models
is well-suited for learning the 
complex mapping between cosmological initial 
conditions and 
proto-halo structures, achieving strong predictive performance \cite{vnet_berger,vnet_bernardini,vnet2024segment,buisman2025differentiablehalomassprediction}.

The V-Net architecture follows an encoder-decoder structure. The contracting path (encoder) captures large-scale contextual features through successive blocks of convolution and  downsampling (via pooling). The expanding path (decoder) performs upsampling and convolution to recover spatial detail and localize structures such as proto-halos. A critical feature of the architecture is the use of skip connections, which directly link corresponding layers in the encoder and decoder \citep{skipconnections}. These connections allow high-resolution feature maps from the encoder to bypass intermediate layers and be combined with the decoder's upsampled features. This mechanism preserves spatial detail and contextual coherence, which is essential for segmenting intricate 3D structures like proto-halos.

In our implementation, the base number of filters for convolutional and transposed-con\-vo\-lu\-tion\-al layers is set to
$M=2P$. Each layer uses a multiple of $M$
filters, ensuring sufficient capacity to allocate at least two feature maps per class. The full architecture for 
$P=5$ output classes is illustrated in figure~\ref{fig:vnetfigure}.
The network consists of 25 convolutional and transposed-convolutional layers in total, with variable kernel sizes of 2, 5, and 8 to allow the model to capture features at multiple spatial scales. Layers are grouped into convolutional blocks, each comprising three layers interleaved with instance normalization \citep{instancenorm} and dropout (with a rate of 40\%). Instance normalization ensures that each input sample is normalized independently of the others in the training batch, improving training stability across diverse samples.
All convolutional and transposed-convolutional layers use the ReLU activation function \citep{relu}, while the output layer uses a softmax activation \citep{NIPS1989_0336dcba}, producing a voxel-wise probability distribution over all output classes. A summary of the architectural components and layer configuration is provided in table~\ref{tab:NN_structure}.

\subsubsection{\textsc{vipr}: a vision transformer
encoder with a CNN decoder}
\begin{figure}
    \centering
    \includegraphics[scale=0.4]{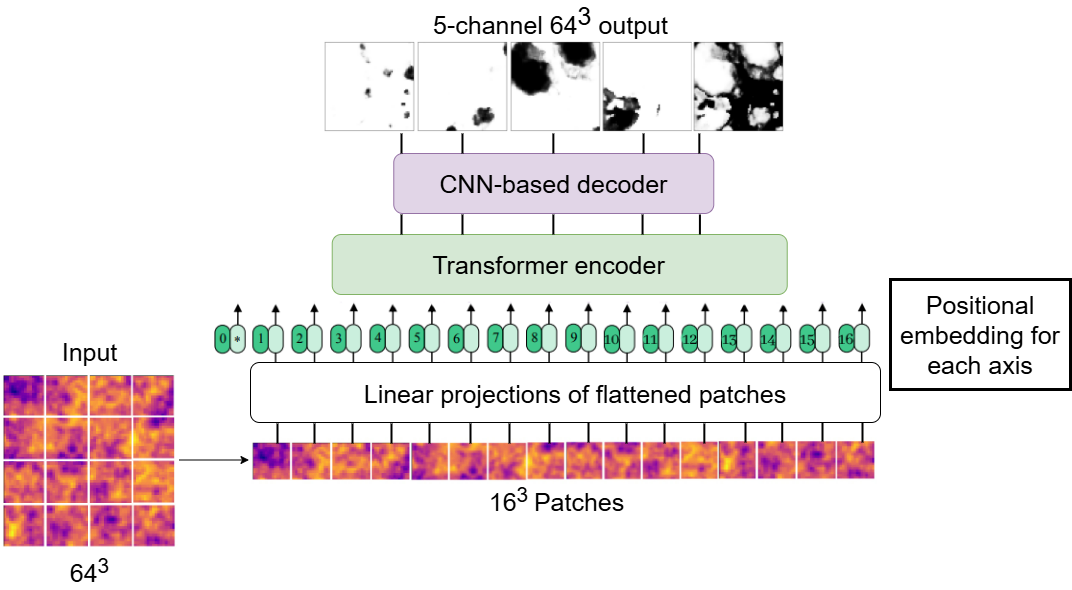}
    \caption{
    Schematic of ViT-based encoder used in \textsc{vipr}.
    The example illustrates the configuration for 5 output classes (corresponding to 4 mass bins plus a non-halo class). The input 3D volume is divided into patches, which are flattened, projected into a 
$K$-dimensional embedding space, and enriched with positional encodings. These embeddings are then processed through a series of transformer blocks. The resulting output is passed to a CNN-based decoder, which reconstructs voxel-wise classification probabilities.
    }
    \label{fig:simplified_vit}
\end{figure}

Our second model, which we call \textsc{vipr} (VIsion transformers for PRoto-halo identification), combines a ViT encoder with a CNN-based decoder,
 and incorporates skip connections between corresponding layers to enhance feature propagation and preserve spatial detail.

Figure~\ref{fig:simplified_vit} illustrates how the transformer operates in practice. Each input sample of size $64^3$ voxels is divided into 64 non-overlapping patches of size $16^3$. These patches are flattened into 1D vectors and mapped into an embedding vector of 
dimension $K=512$ by applying a learned linear transformation. Because transformers are permutation-invariant, we add positional encodings to each embedding.
The encoder takes the 64 embeddings of the patches (with added positional encodings) and processes them through
a stack of transformer blocks. Each block consists of two main components: a multi-head self-attention layer, which enables the model to capture relationships between patches, and a feed-forward network, which non-linearly transforms each patch embedding after incorporating contextual information from the attention layer.
The output is subsequently reshaped into a structured 3D feature map and passed to the CNN-based decoder, which progressively upsamples it back to the original resolution.

\begin{figure}
    \centering
    \includegraphics[scale=0.4]{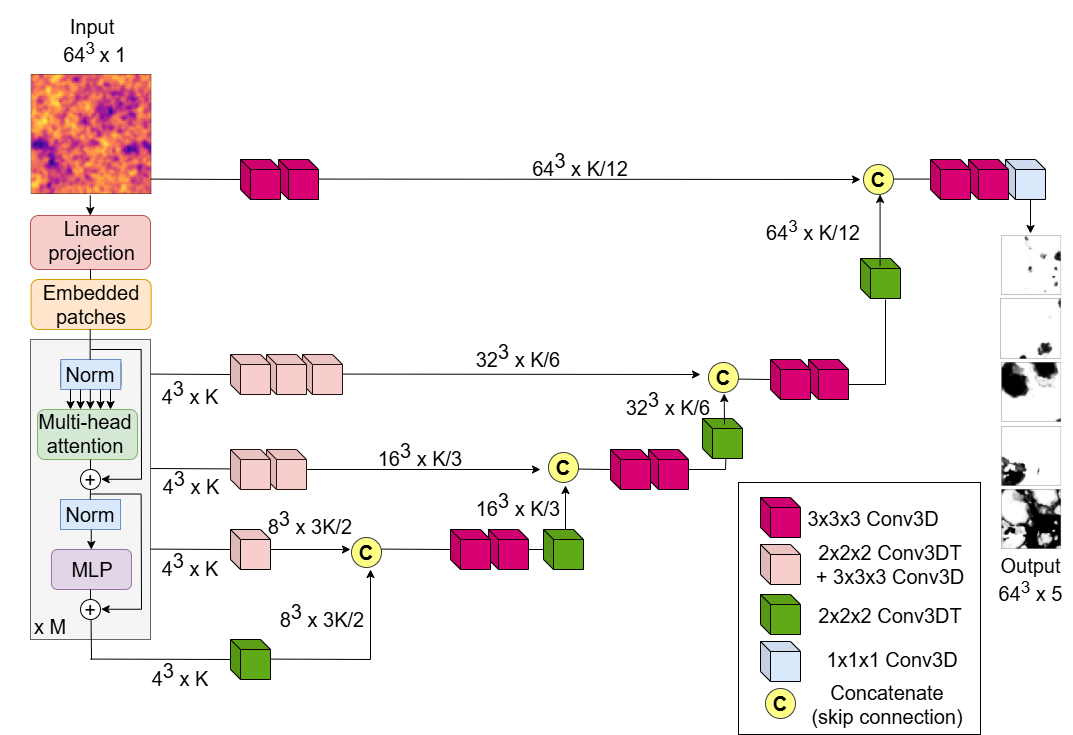}
    \caption{Full UNETR architecture used in \textsc{vipr}. 
    The base number of filters is set to $M=2P$, where 
$P$ is the number of output classes. The ViT encoder uses $M$ attention heads and an embedding size of $K=512$.
In the CNN-based decoder, each convolutional block includes two instance normalization layers. Skip connections from selected transformer layers are integrated into the decoder to retain spatial context and support accurate volumetric segmentation.}
    \label{fig:full_vit}
\end{figure}
The complete architecture of the model is shown in Figure~\ref{fig:full_vit}. Similar to \textsc{copra}, it takes the 3D initial density field as input and outputs a probability mask for proto-halo classification. The ViT encoder captures long-range dependencies and global spatial patterns through self-attention, while the CNN decoder upsamples these representations to generate the final segmentation mask.
The model is implemented using the \textsc{self-attention-cv} library,\footnote{https://github.com/The-AI-Summer/self-attention-cv} which offers a configurable framework for UNETR architectures. We set the number of ViT attention heads and base filters to $M=2P$, matching the \textsc{copra} configuration. The transformer encoder employs the GELU activation function \cite{hendrycks2016gelu}, while all convolutional and transposed-convolutional layers in the decoder use ReLU. As in \textsc{copra}, each convolutional layer is followed by two Instance Normalization layers. The output layer applies softmax to produce class probabilities. Dropout was omitted as training proved more stable without it.
Compared to \textsc{copra}, this network is more complex, with $M=2P$ attention heads and 27 convolutional or transposed-convolutional layers. Its architecture is summarized in Table~\ref{tab:NN_structure}.

\begin{table}
\centering
\begin{tabular}{@{}lcccccccccccc}
\hline
Model & Levels & \multicolumn{5}{c}{ \# Conv} & \multicolumn{5}{c}{\quad\# TConv}& ViT Heads\\
& & $1^3$ & $2^3$ & $3^3$ & $5^3$ & $8^3$ &       & &$2^3$ &  $8^3$ & &\\ 
\hline
\textsc{copra} & 4 & 1& 3  & 0 & 12 & 5 &       &  &3 &  5 & &-- \\
\textsc{vipr}  & 4 & 1 & 0 & 16 &0 & 0 &        & & 10 &  0 & &$M=2P$  \\
\hline
\end{tabular}

    \caption{Overview of the architecture of our deep neural network models. 
    The table summarizes key design parameters, including kernel sizes, the number of convolutional and transposed-convolutional layers, and the number of ViT heads used in the \textsc{vipr} model. A “level” refers to a convolutional block that is linked to a corresponding transposed-convolutional layer through a skip connection.}
    \label{tab:NN_structure}
\end{table}

\subsection{Training and optimization}
For both models, we adopt the Dice loss function \cite{dice}, a widely used choice for image segmentation tasks due to its effectiveness in handling class imbalance and its direct optimization of overlap metrics \cite{surveydice, gendiceovrlp}.
 In the multi-class setting, let $\mathbf{p}_c$ and $\mathbf{g}_c$ denote the soft probabilistic prediction and the binary ground truth mask for class $c$, respectively. The Dice coefficient for each class is defined as
\begin{equation}
\mathrm{Dice}_c = \frac{2\, |\mathbf{p}_c \cdot \mathbf{g}_c|}{|\mathbf{p}_c|^2 + |\mathbf{g}_c|^2}\;,
\end{equation}
ranging from 0 (no overlap) to 1 (perfect overlap).
To obtain a global measure of performance across all $P+1$ classes, the Dice scores can be averaged either uniformly (unweighted) or in a weighted fashion to reflect class prevalence. The corresponding loss function is
\begin{equation}
\mathrm{Loss} = 1 - \sum_{c=1}^{P+1} w_c\,\mathrm{Dice}_c\;,
\end{equation}
where the weights $w_c$ may be chosen to counteract class imbalance (e.g., using inverse class frequency).
The loss is minimized when the predicted probability maps best align with the true segmentation masks.

We train each model for 180 epochs using three out of four N-body simulations, comprising a total of 10,125 input-output samples. The fourth simulation is held out entirely for testing purposes and is not used during training or validation.

To facilitate training, we normalize the density field
$\delta$ to the range $[0,1]$, applying the same transformation to both training and testing inputs. This rescaling places the input values in a numerically stable range, promoting faster convergence and improved training stability.

Training is performed on an NVIDIA A40 GPU with 48 GB of memory. During each epoch, the model iterates once through the entire training dataset in mini-batches. We allocate 85\% of the training data for learning and 15\% for validation,  and we monitor the validation accuracy after each epoch to assess the model’s generalization performance.
Model checkpoints are used to retain only the version that achieves the highest validation accuracy, regardless of the epoch at which it occurs.

We fine-tune the hyperparameters separately for each model architecture.
For the CNN-based model (\textsc{copra}), we use the Adam optimizer \cite{adam} with a learning rate of 
$10^{-3}$ and a batch size of 64. For the ViT-based model (\textsc{vipr}), we use the AdamW optimizer \cite{adamw}, a variant designed for weight decay regularization, with a learning rate of 
$10^{-4}$ and a batch size of 4.

\subsection{Full-volume reconstruction}
\label{sec:scores}
To reconstruct predictions for the full test volume, we must reconcile the fact that many voxels appear in multiple mini-cubes and thus receive multiple predictions.
We aggregate these overlapping predictions using a maximum-confidence strategy: for each voxel, we retain the full soft score vector from the mini-cube where the network's output exhibits the highest confidence, defined as the maximum class probability across all overlapping predictions. This approach preserves the probabilistic output while prioritizing the most decisive estimate.
As an alternative, we also experimented with averaging the soft score vectors from all mini-cubes containing the voxel. While both methods yield similar predictions near the centers of proto-halos, the maximum-confidence approach tends to produce sharper boundaries --- especially in regions where classification uncertainty is typically higher --- at the cost of a potentially increased sensitivity to outliers.

We observe a striking difference in the score distributions produced by the two models: while
\textsc{copra} exhibits a nearly bimodal distribution, with scores strongly peaked near 0 and 1, \textsc{vipr}'s output displays a richer range of intermediate values, indicating more nuanced probabilistic predictions. This behavior reflects the distinct inductive biases of the two architectures: CNNs are built to capture localized patterns and are more prone to producing sharp activations, while ViTs leverage global self-attention, enabling them to account for broader spatial context and thus express more nuanced and distributed confidence. These differences impact not only the visual character of the output masks but also the interpretability and thresholding strategies (see section~\ref{sec:optimal_thresholds}). In particular, ViTs' smoother score maps may be advantageous in capturing transitional regions or expressing uncertainty, whereas CNNs' binary-like outputs are easier to binarize but may be more brittle near class boundaries. 

\begin{figure}
    \centering
    \includegraphics[scale=0.2]{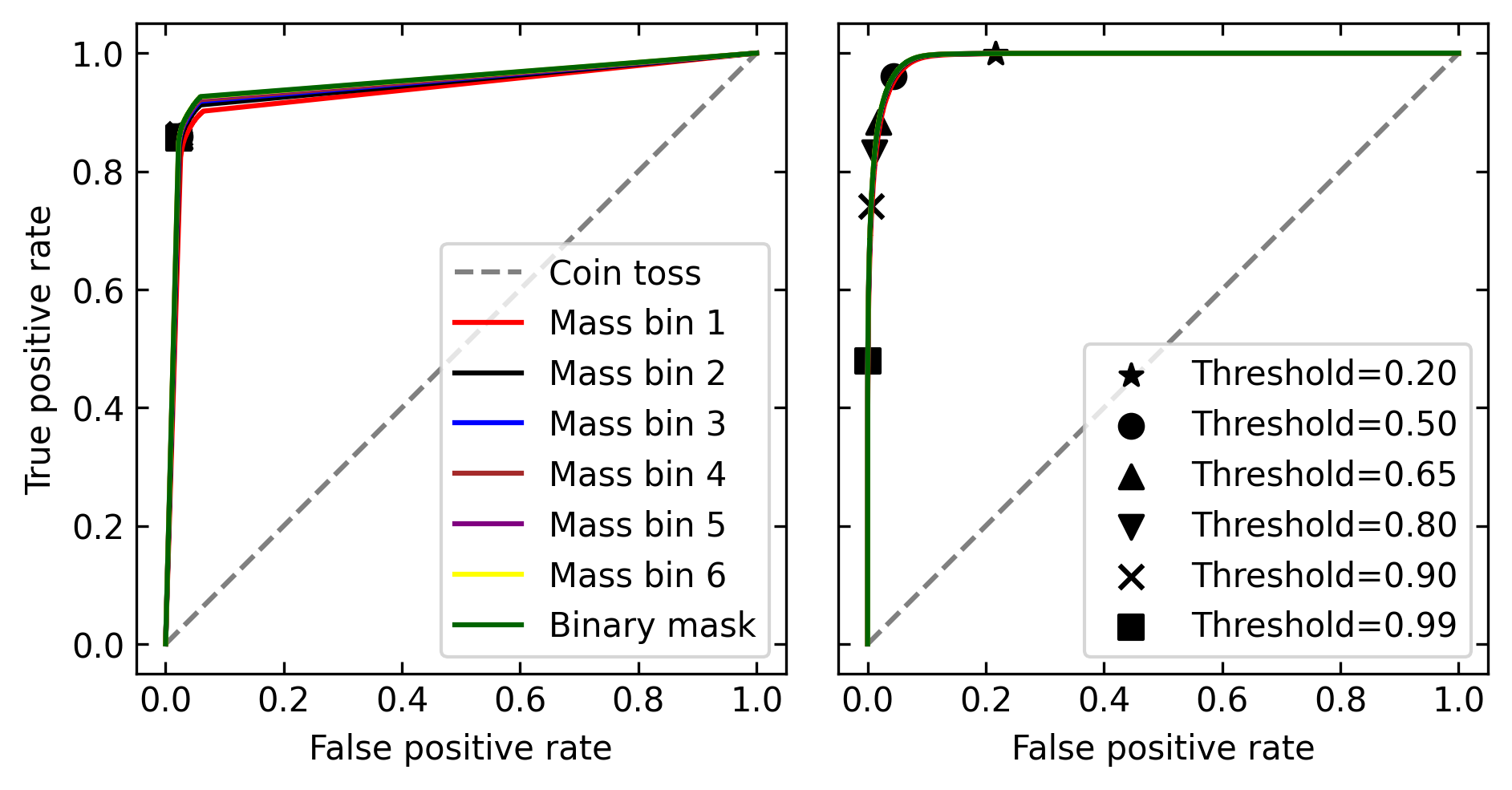}
    \caption{
    Receiver Operating Characteristic (ROC) curves for the best-performing CNN-based model (\textsc{copra}, left) and ViT-based model (\textsc{vipr}, right), each trained with 7 output classes. The curves, computed on the validation set, show the binary classification performance for all classes; due to the high accuracy, the curves for different classes nearly overlap. Symbols indicate selected operating points for different classification thresholds, shown only for the non-halo (background) class. In the left panel, the symbols nearly coincide due to the sharpness of \textsc{copra}'s score distribution.}
    \label{fig:rocs}
\end{figure}

\subsection{Optimal classification thresholds}
\label{sec:optimal_thresholds}

To convert the probabilistic outputs of our models into discrete class predictions, a classification threshold must be applied. A voxel is classified as ``positive'' if its predicted probability exceeds the threshold. Comparing this classification to the ground truth determines whether each voxel is a true positive (TP, correctly classified as halo), false positive (FP, incorrectly classified as halo), true negative (TN, correctly classified as non-halo), or false negative (FN, incorrectly classified as non-halo). In segmentation tasks, the choice of the classification threshold directly impacts key evaluation metrics such as purity and completeness. 
Purity (also known as precision) quantifies the fraction of predicted positives that are correct:
$\mathrm{Purity} = \frac{\mathrm{TP}}{\mathrm{TP}+\mathrm{FP}}$,
while,
completeness (also called recall or sensitivity) measures the fraction of actual positives that are successfully identified:
$\mathrm{Completeness} = \mathrm{TPR} = \frac{\mathrm{TP}}{\mathrm{TP} + \mathrm{FN}}$.

Although assigning to each voxel the class with the highest predicted probability (i.e., argmax) is a common strategy, per-class thresholding offers a more flexible alternative, particularly valuable in cases of class imbalance or differing calibration between output classes. By tuning thresholds independently, one can prioritize completeness for rare classes or purity for dominant ones, tailoring the output to specific scientific goals. The Receiver Operating Characteristic (ROC) curve provides a principled framework to guide this threshold selection. It plots the TPR against the false positive rate
$\mathrm{FPR} = \frac{\mathrm{FP}}{\mathrm{FP} + \mathrm{TN}}$,
as the classification threshold varies.
Figure~\ref{fig:rocs} shows the ROC curves, computed on the validation data,
for the models trained with 7 output classes. These curves form the basis for selecting optimal classification thresholds, which can be determined using various criteria. A straightforward and widely used method
is to maximize the F1 score for each class, defined as the harmonic mean of purity and completeness: 
F1$=\frac{2\,\mathrm{TP}}{2\,\mathrm{TP}+\mathrm{FP}+\mathrm{FN}}$.
This metric is mathematically equivalent to the Dice coefficient, provided the latter is computed on hard (binarized) classifications.
Another approach involves minimizing the Euclidean distance to the ideal point on the ROC curve (i.e., the perfect classifier with $\mathrm{FPR} = 0$ and $\mathrm{TPR} = 1$), which identifies the threshold yielding the best compromise between sensitivity and specificity. 
Alternatively, one may maximize Youden’s $J$ statistic, defined as $J=\mathrm{TPR}-\mathrm{FPR}$, to capture the overall discriminative power. 
Geometrically, this corresponds to the point on the ROC curve that lies farthest from the diagonal (chance line) in the vertical direction.
In practice, the optimal choice depends on the application: tasks prioritizing completeness (e.g., detecting all relevant voxels) may tolerate more false positives, while those requiring high purity may favor stricter thresholds.

\begin{table}
    \centering
     \begin{tabular}{c@{\hspace{14pt}}@{\hspace{14pt}}ccc@{\hspace{14pt}}@{\hspace{14pt}}ccc}
    \hline
     & \multicolumn{3}{c}{\textsc{copra}} & \multicolumn{3}{c}{\textsc{vipr}} \\
     Halo mass bin & $\max \text{F1}$ & $\min D(0,1)$ & $\max J$ & $\max\text{F1}$ & $\min D(0,1)$ & $\max J$ \\
    \hline
    1 & 0.99 & 0.99 & 0.99 & 0.53 & 0.53 & 0.53 \\
    2 & 0.99 & 0.99 & 0.99 & 0.55 & 0.55 & 0.55 \\
    3 & 0.99 & 0.99 & 0.99 & 0.55 & 0.55 & 0.56 \\
    4 & 0.99 & 0.99 & 0.99 & 0.52 & 0.52 & 0.52 \\
    5 & 0.99 & 0.99 & 0.99 & 0.55 & 0.55 & 0.56 \\
    6 & 0.92 & 0.91 & 0.90 & 0.59 & 0.60 & 0.60 \\
    Non-halo & 0.99 & 0.99 & 0.99 & 0.62 & 0.61 & 0.62 \\
    \hline
    \end{tabular}
    \caption{
Optimal classification thresholds per output class, selected according to three different criteria: maximum F1 score (equivalent to the maximum Dice coefficient for hard classifications), minimum Euclidean distance to the ideal point $(0,1)$ on the ROC curve, and maximum Youden's J statistic. All metrics are computed on the validation dataset using soft probability outputs from the models.}
    \label{tab:thresholds_per_class}
\end{table} 

Table~\ref{tab:thresholds_per_class} reports the resulting optimal thresholds per class, obtained using these criteria on the validation set. Despite the different definitions, the selected thresholds for each model are nearly identical across criteria, indicating consistent calibration of the predicted scores.
Notably, the optimal thresholds for \textsc{copra} cluster tightly near 0.99 for most classes, dropping slightly to approximately $0.90$ in the highest-mass bin. In contrast, \textsc{vipr} yields significantly lower thresholds, typically ranging between 0.53 and 0.62. This difference reflects the distinct behavior of the two architectures: \textsc{copra} produces more sharply peaked confidence scores, while \textsc{vipr} shows a broader, more distributed confidence spectrum, as also visualized in figure~\ref{fig:rocs} and discussed in section~\ref{sec:scores}.

\section{Results}
\label{sec:results}

\begin{figure}
    \centering
    \includegraphics[scale=0.075]{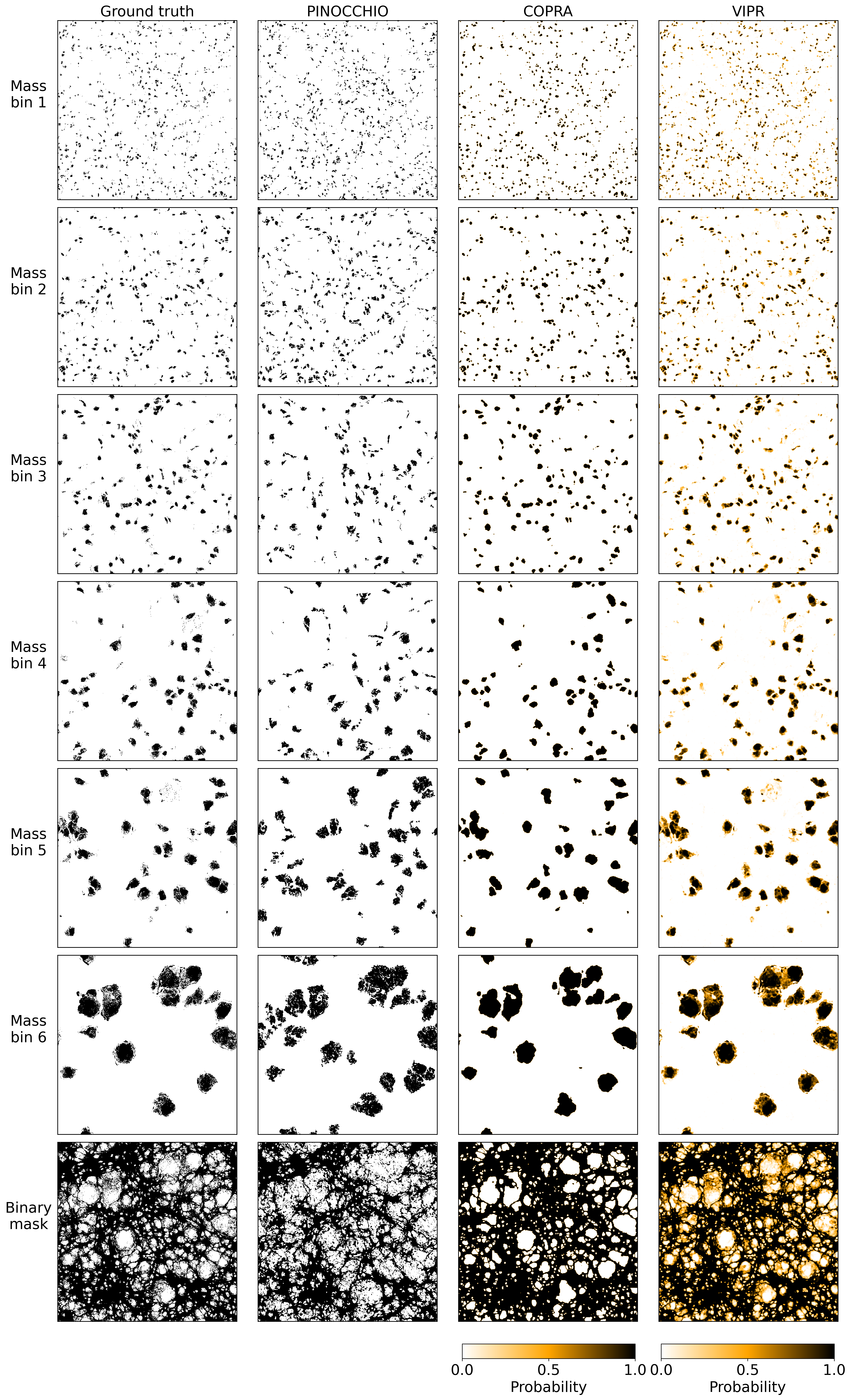}
    \caption[Short caption]{Single-voxel slice through the test simulation box, comparing proto-halo predictions across methods.}
    \label{fig:visual_res}
\end{figure}

We begin our analysis by visually comparing the predictions produced by our deep learning models to the ground truth. Figure~\ref{fig:visual_res} shows a representative example from the held-out test simulation, displaying a single-voxel thick slice of the 3D volume. The comparison includes the true proto-halo segmentation, alongside predictions from our CNN-based model (\textsc{copra}), ViT-based model (\textsc{vipr}), and, for reference, the perturbation-theory-based model \textsc{pinocchio} \cite{pinocchio,Monaco_2002}. 
\textsc{pinocchio} (v. 4.1.2) was run with the same initial conditions as our simulations, employing second-order LPT for group finding and third-order LPT for displacements, the latter being irrelevant for the present analysis.
Each panel highlights the voxel-wise classification into the different mass bins of dataset 3, as well as the non-halo (background) class. This example offers a first impression of the segmentation capabilities of each method, and allows for a visual inspection of similarities and discrepancies in predicting proto-halo morphology and spatial extent.

Although all methods produce qualitatively similar outputs at first glance, a closer inspection reveals clear differences in fidelity with respect to the ground truth. Among the models, \textsc{copra} and \textsc{vipr} yield the most accurate and spatially coherent predictions, closely matching the true proto-halo regions.
In contrast, \textsc{pinocchio} shows the largest discrepancies: it overestimates the mass fraction in halos --- clearly visible in the panel corresponding to the non-halo class (denoted as the binary mask) --- and occasionally generates spurious structures (see, for example, the bottom-right corner of the figure for the high-mass bin).
To better highlight these differences, figure~\ref{fig:visual_res_color} in appendix~\ref{appendix:voxel_agreement} reproduces the same slice using a color-coded scheme that distinguishes correctly and incorrectly classified voxels for each method. This visualization makes the strengths and limitations of each approach more evident at the voxel level.

\subsection{Score distributions and confidence calibration}

\begin{figure}
    \centering
    \includegraphics[width=0.8\linewidth]{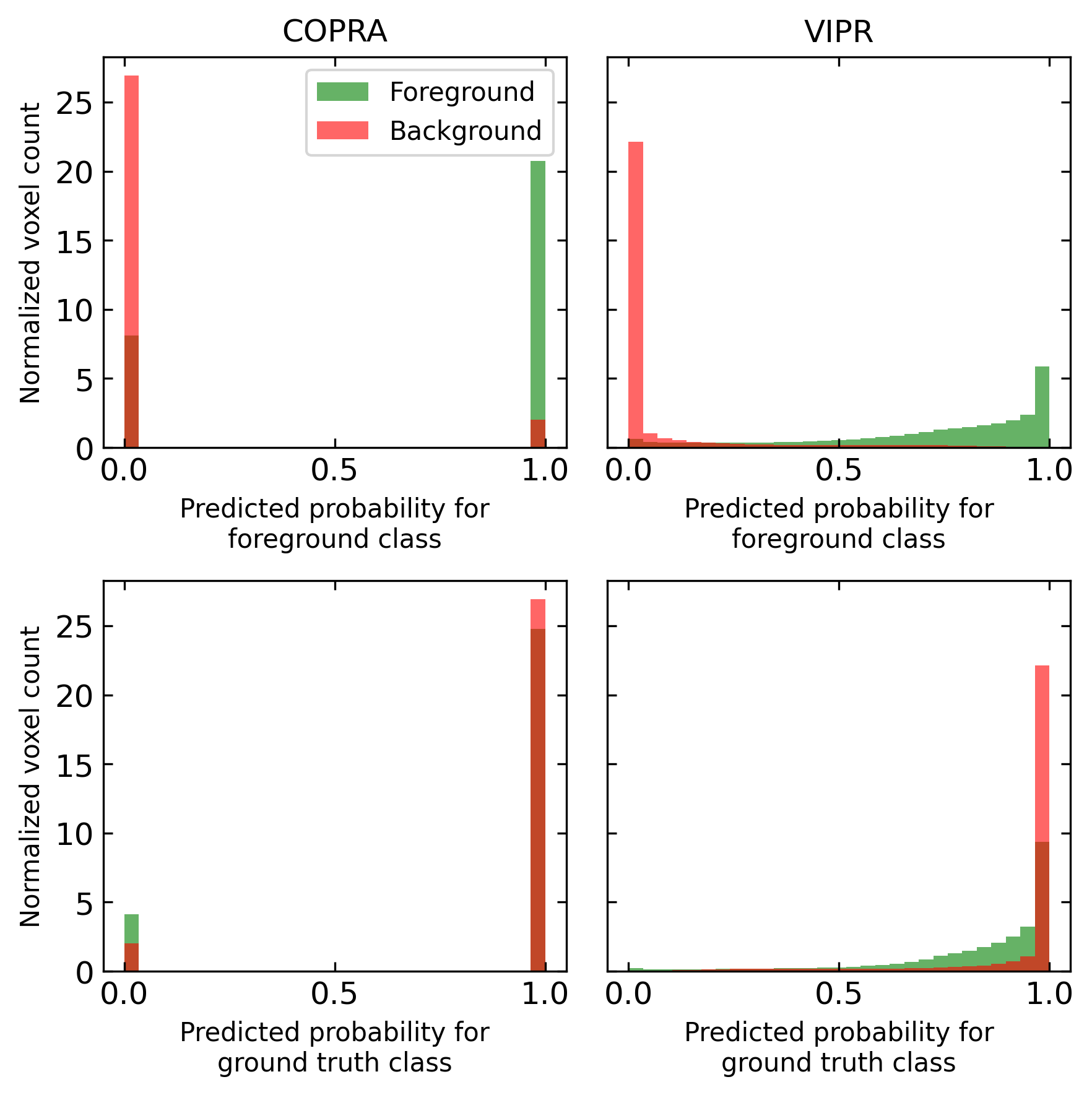}
    \caption{Distributions of predicted voxel scores for \textsc{copra} and \textsc{vipr}, computed from the test simulation. Top panels: distributions of the predicted probability for the aggregated foreground class, obtained by summing the probabilities assigned to all halo classes (or equivalently, one minus the probability of the non-halo class). Histograms are shown separately for voxels truly belonging to the foreground and background, with each distribution normalized to unit area to emphasize the separation between the two regions.
Bottom panels: distributions of the predicted probability assigned to the ground truth class of each voxel, again presented separately for background and aggregated foreground voxels.}
    \label{fig:score_hist}
\end{figure}

To move beyond visual inspection, we next analyze
how the models assign probabilities to voxels and how confidently they distinguish halo from non-halo regions. This analysis is carried out on the entire held-out test set and focuses on models trained on dataset 3 ($P=7$ output classes). We characterize the predictions by examining the distribution of voxel-level scores using two complementary approaches.

The top panels of figure~\ref{fig:score_hist} show the distribution of the predicted probability for the aggregated foreground class, calculated as the sum of probabilities assigned to all halo classes (or equivalently, one minus the probability of the non-halo class). These histograms are computed from the test simulation and are shown separately for voxels truly belonging to the foreground and background. They highlight the models’ ability to discriminate collapsed from non-collapsed regions.
\textsc{copra} produces an almost bimodal distribution, 
with probabilities clustered near 0 or 1, indicating extreme confidence.
However, this also results in 24.3\% of true foreground voxels being misclassified with near-zero probabilities, and 7.1\% of background voxels receiving probabilities close to 1, revealing overconfident mistakes. In contrast, 
\textsc{vipr}, yields
broader score distributions for both true foreground and background voxels. Despite this spread, the two distributions remain well separated, reflecting reliable discrimination. Importantly, voxels with high-confidence wrong classifications are rare, suggesting that \textsc{vipr}’s confidence, even when dispersed, is generally well calibrated.

The bottom panels present the distributions of predicted probabilities conditioned on the ground truth class,
where for each voxel we use the probability that the network assigns to its actual class (one of the seven possible ones).
Again, voxels are grouped into foreground and background. This representation directly reveals how confident the models are  in making
their predictions.
For \textsc{copra}, the distribution is again sharply bimodal,
with a secondary peak near zero corresponding to misclassified voxels, reflecting its tendency to be either very correct or very wrong.
\textsc{vipr}, by contrast, shows a narrow distribution for the background voxels (indicating strong confidence where predictions are correct) and a wider spread for foreground voxels. The presence of a few low-confidence predictions is natural, but importantly, high-confidence misclassifications are almost absent, indicating robust reliability in its probabilistic outputs.

Overall, these results demonstrate that while \textsc{copra} exhibits overconfidence and a higher incidence of confident errors, \textsc{vipr} provides better-calibrated probabilistic outputs that maintain accuracy without sacrificing reliable uncertainty estimates.

\subsection{Model evaluation and performance metrics}
Finally, we quantify the segmentation performance 
of our models on the held-out test set to assess generalization.
To this end, we use the ROC curves and their associated Area Under the Curve (AUC) scores. 
These metrics provide a threshold-independent view of classification performance, which is particularly useful given the probabilistic nature of our model outputs.
Table~\ref{tab:auc_dataset1} summarizes the AUC scores across all datasets.
They range from 0.91 to 0.95 for
\textsc{copra},
and from 0.98 to 0.99 for
\textsc{vipr}.
Since AUC scores above 0.8 are generally considered indicative of good classification performance, and scores above 0.9 reflect highly discriminative models, we conclude that both networks yield accurate predictions --- especially \textsc{vipr}, which consistently achieves superior AUC values across all datasets and classes.

\begin{table}[]
    \centering
    \begin{tabular}{cccccccccc}
    \hline
    Model & Dataset & \multicolumn{6}{c}{Halo mass bin} & Binary mask\\
           &  & 1 & 2 & 3 & 4 & 5 & 6 &\\
             \hline
     & 1 & 0.91 & 0.92 & 0.92 & 0.93 &- & - & 0.93\\   
     \textsc{copra} & 2 & 0.93 & 0.93 & 0.94 & 0.94 & 0.94 & - & 0.94\\ 
     & 3 & 0.93 & 0.94 & 0.94 & 0.95 & 0.95 & 0.95 & 0.95\\ 
    \hline
    & 1 & 0.98 & 0.98 & 0.98 & 0.98 &- & - & 0.98\\   
     \textsc{vipr} & 2 & 0.99 & 0.99 & 0.99 & 0.99 & 0.99 & - & 0.99\\ 
     & 3 & 0.99 & 0.99 & 0.99 & 0.99 & 0.99 & 0.99 & 0.99\\ 
    \hline
    \end{tabular}
    \caption{The Area Under the Curve (AUC) score for each class (i.e. halo mass bin), computed from the ROC curves of the predictions on the test dataset. These scores quantify the model's predictive ability for the different classes across different classification thresholds.}
    \label{tab:auc_dataset1}
\end{table}

To complement the threshold-independent AUC scores, table~\ref{tab:val_accs} displays the training, validation, and test accuracies of the best-performing models across all datasets. 
These values are computed by converting the soft probabilistic outputs into discrete class predictions.
For each voxel, we select the class with the highest predicted probability (the argmax rule),\footnote{Since no voxel has a maximum class probability below 0.5, this is effectively equivalent to applying a uniform classification threshold of 0.5.} and then compare the resulting labels to the ground truth. 
The accuracy is defined as the fraction of voxels that are correctly classified.
We adopt the argmax criterion because the optimized thresholds presented in section~\ref{sec:optimal_thresholds} are derived from the validation phase and therefore cannot be applied during training.
\textsc{copra} achieves validation accuracies around 85\%, while \textsc{vipr} consistently exceeds 90\%, demonstrating superior performance. In all cases, the training, validation, and test accuracies are closely aligned, which suggests that the models generalize well to unseen data and do not suffer from overfitting.

For both models, increasing the learning capacity --- by raising the number of base filters or the number of attention heads in the ViT encoder --- generally improves performance, particularly when accompanied by a higher number of output channels. However, we also observe that increasing model complexity without adjusting the output dimensionality yields diminishing returns and may lead to overfitting. 

From this point onward, we restrict our analysis to models trained on dataset 3 (i.e., with 7 output classes), which represent the best-performing configurations. 

To ensure consistency with the training objective, 
we adopt for all subsequent analyses
the classification thresholds that maximize the Dice score on the validation set (listed in table~\ref{tab:thresholds_per_class}). 
Because these thresholds can leave a small fraction of voxels unclassified (if no class probability exceeds the respective threshold), we apply a fallback strategy: these voxels --- only $0.003\%$ for \textsc{copra} and $0.06\%$ for \textsc{vipr} --- are assigned to the non-halo class.\footnote{Alternative fallback strategies include assigning unclassified voxels to the class with the highest probability (i.e., an argmax fallback), redistributing them based on prior class frequencies, or omitting them altogether from the final mask. However, such choices may either bias the class distribution or reduce interpretability in a physical context.}
This approach is physically motivated, as low-confidence voxels are unlikely to belong to well-defined proto-halos. It also preserves the overall consistency of the segmentation while avoiding spurious high-mass classifications driven by uncertain predictions.

Using these thresholds, we evaluate the model performance at the voxel level on the test simulation. Table~\ref{tab:test_accs} presents the per-class test accuracy,
confirming that overall accuracy is not dominated by the abundant non-halo class but reflects consistent performance across all halo mass bins. For \textsc{vipr}, the test accuracy remains remarkably stable at approximately 97.5\% for all halo classes, with a slight drop for the background (85.3\%). In contrast, \textsc{copra} shows a gradual decline in accuracy with increasing halo mass, from 94.3\% in the lowest mass bin to 88.0\% in the highest. The background accuracy for \textsc{copra} is similar to that of \textsc{vipr} (85.2\%). These results highlight the robustness of \textsc{vipr} across all mass classes, whereas \textsc{copra} exhibits a mild sensitivity to halo mass.

\begin{table}
    \small
    \centering
    \begin{tabular}{c@{\hspace{14pt}}@{\hspace{14pt}}ccc@{\hspace{14pt}}@{\hspace{14pt}}ccc}

    \hline
     & \multicolumn{3}{c}{\textsc{copra}} & \multicolumn{3}{c}{\textsc{vipr}} \\
   Dataset & Training & Validation & Test & Training & Validation & Test \\
    \hline
    1 & 84.7 & 83.8 & 83.6 & 91.2 & 90.8 & 90.6 \\
    2 & 85.2 & 84.6 & 83.8 & 92.3 & 91.5 & 90.7 \\
    3 & 86.3 & 85.2 & 84.3 & 92.7 & 92.4 & 91.2 \\
    \hline
    \end{tabular}
    \caption{Training, validation, and test accuracies (in percent) of the neural network models with optimized hyperparameters, trained separately on each of the three datasets.}
    \label{tab:val_accs}
\end{table}

\begin{table}
    \centering
    \begin{tabular}{c c c c c c c c}
    \hline 
Model & \multicolumn{6}{c}{Halo mass bin}& Non-halo \\
& 1 & 2 & 3 &4 &5 & 6 & \\
\hline
\textsc{copra} & 94.3 & 94.2 & 92.1 & 92.0 & 88.1 & 88.0 & 85.2\\
\textsc{vipr} & 97.4 & 97.7 & 97.7 & 97.6 & 97.3 & 97.9 & 85.3\\
\hline    
    \end{tabular}
    \caption{Per-class test accuracies (in percent) evaluated on the held-out test simulation using dataset 3. These accuracies are computed using the otpimal classification thresholds.}
    \label{tab:test_accs}
\end{table}

\begin{table}[]
    \centering
    \begin{tabular}{c@{\hspace{14pt}}@{\hspace{14pt}}ccc@{\hspace{14pt}}@{\hspace{14pt}}ccc}

\hline
 & \multicolumn{3}{c}{\textsc{copra}} & \multicolumn{3}{c}{\textsc{vipr}} \\
Halo mass bin & TPR & FPR & Dice & TPR & FPR & Dice \\
\hline
1 & 0.75 & 0.02 & 0.69 & 0.73 & 0.01 & 0.79 \\
2 & 0.82 & 0.02 & 0.76 & 0.87 & 0.01 & 0.82 \\
3 & 0.86 & 0.02 & 0.80 & 0.88 & 0.01 & 0.84 \\
4 & 0.87 & 0.02 & 0.82 & 0.88 & 0.01 & 0.84 \\
5 & 0.88 & 0.02 & 0.82 & 0.89 & 0.02 & 0.84 \\
6 & 0.89 & 0.02 & 0.84 & 0.92 & 0.02 & 0.88 \\
Non-halo & 0.93 & 0.27 & 0.89 & 0.91 & 0.15 & 0.91 \\
\hline
\end{tabular}
\caption{
Evaluation metrics for each output class on the test dataset, computed using the optimal thresholds determined by maximizing the Dice score on the validation set.}
\label{tab:test_metrics_dice_threshold}
\end{table}

Finally, table~\ref{tab:test_metrics_dice_threshold} reports the per-class TPR, FPR, and Dice coefficients. These additional metrics confirm that thresholds selected on validation data generalize well, preserving the segmentation quality for unseen data.

\subsection{Mass fraction recovery across halo classes}

As a complementary test, we evaluate the models’ ability to recover the total halo mass in each bin across the entire test volume. This is conceptually related to the halo mass function but does not require fragmenting the predicted segmentation into individual halos.

\begin{figure}
    \centering
    \includegraphics[scale=0.15]{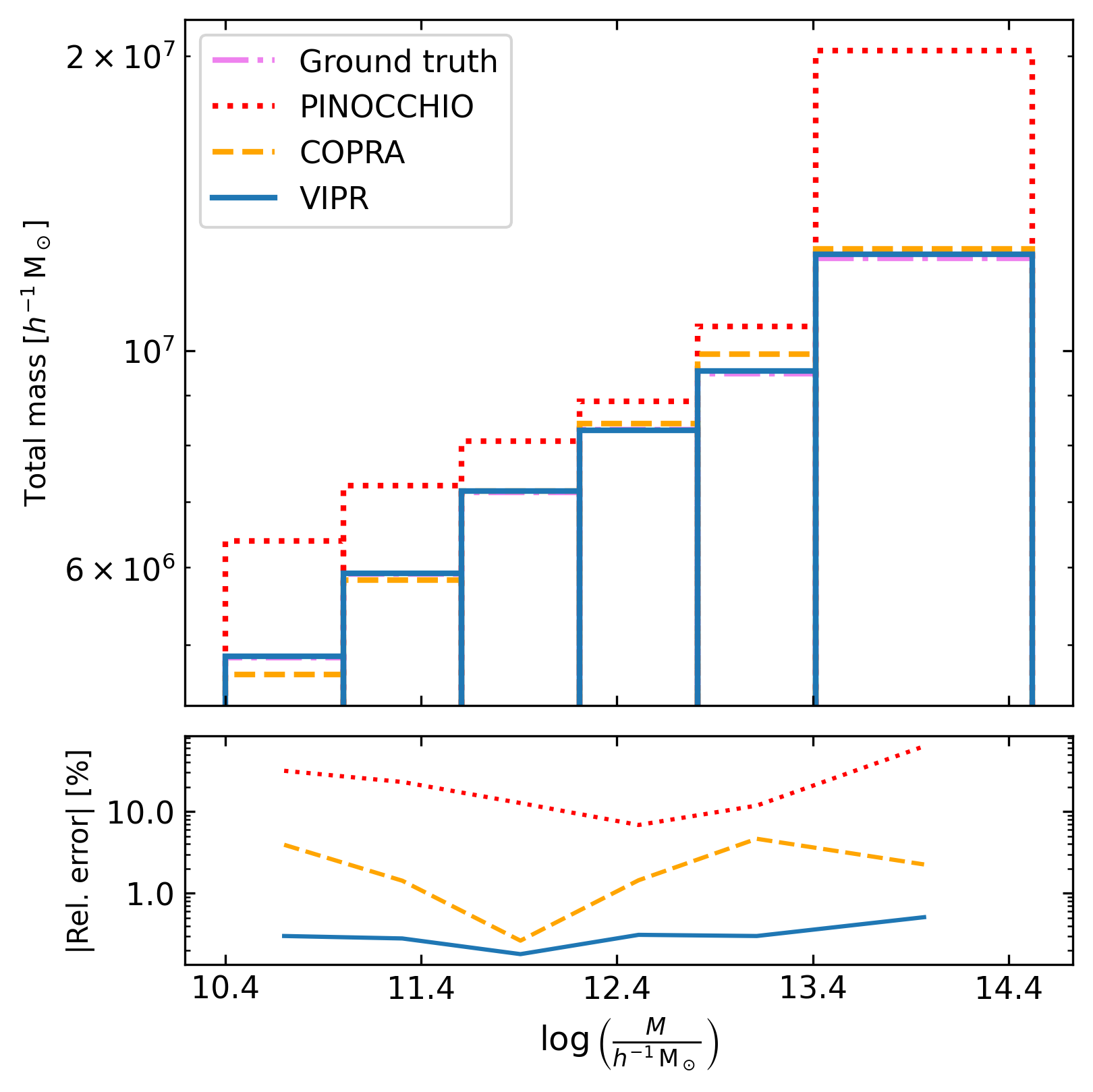}
    \caption{
Total mass in halos predicted by the models across different halo mass bins, compared to the ground truth from the test simulation at $z=0$. 
The bottom panel shows the absolute relative deviation from the ground truth for each method.}
\label{fig:histogram}
\end{figure}

Figure~\ref{fig:histogram} compares the predicted mass distributions to the ground truth, alongside results from the perturbation-theory-based \textsc{pinocchio} model for reference.
\textsc{vipr} clearly outperforms all other methods, achieving highly consistent sub-percent accuracy, with relative errors ranging from just 0.2\% to 0.9\% across all mass bins.
\textsc{copra} exhibits relative errors between 0.3\% and 5\%, but its predictions are systematically less accurate than those of \textsc{vipr}. In contrast, \textsc{pinocchio} shows significantly larger deviations: it overestimates the total mass by about 30\% at low masses, shows a minimum error of 10\% at intermediate scales, and reaches errors exceeding 60\% in the highest-mass bin. 
\textsc{pinocchio} is generally found to be in good agreement with $N$-body simulations, reproducing halo mass functions at the $\sim$10–20\% level when compared under consistent conditions \citep[e.g.][]{Monaco_2002,Monaco_2013,Munari+2017,Euclid_2023}. The larger discrepancies we observe here stem mainly from the fact that \textsc{pinocchio} is calibrated against friends-of-friends (FOF) halos (with linking length $b=0.2$),
whereas our simulations adopt a spherical overdensity (SO) criterion with $\Delta=200$ relative to the mean density. Since FOF halos are generally more massive than SO halos, systematic offsets between the two catalogs are expected. 
We have verified that the halos in our N-body simulations are consistent with standard fitting functions for SO-based mass functions \citep[e.g.][]{Tinker_2008}, and that the \textsc{pinocchio} mass function itself shows good agreement with FOF-based fits \citep[e.g.][]{Warren_2006} at low masses. This confirms that the discrepancies we observe mainly reflect differences in halo definition rather than shortcomings of either method.

\subsection{Object-level performance assessment}

While our models demonstrate strong performance when evaluated on the entire test simulation volume, it is essential to assess whether this accuracy persists at the level of individual proto-halos. 
In this work, we do not address the related but distinct problem of instance segmentation --- that is, the task of identifying and separating individual proto-halo regions.
Instead, we assess the voxel-wise classification accuracy within specific halo structures drawn from the ground truth.

To this end, we select six representative halos of different masses and, for each of them, we examine how well the models predict the corresponding proto-halo structures. The first four columns of table~\ref{tab:objects_errors} list the label assigned to each halo, the precise mass value, the number of simulation particles it contains, the Lagrangian radius (computed from the mass assuming a spherical proto-halo), and the maximum Lagrangian extent --- defined as the greatest distance from any proto-halo voxel to its center of mass.
The top entry, labeled H1, corresponds to the most massive halo in the test box. The next two (H2 and H3) refer to individual group-sized halos, while the final three (H4, H5, and H6) represent galaxy-sized halos of progressively smaller mass, whose proto-halo regions become increasingly challenging to capture accurately given the fixed mass resolution of the simulation.
The halos were selected at random from the test set, with the exception of H-1, which was deliberately included as the most massive halo in the box in order to illustrate model performance at the extreme high-mass end.

\begin{figure}
    \centering
    \includegraphics[width=0.76\linewidth]{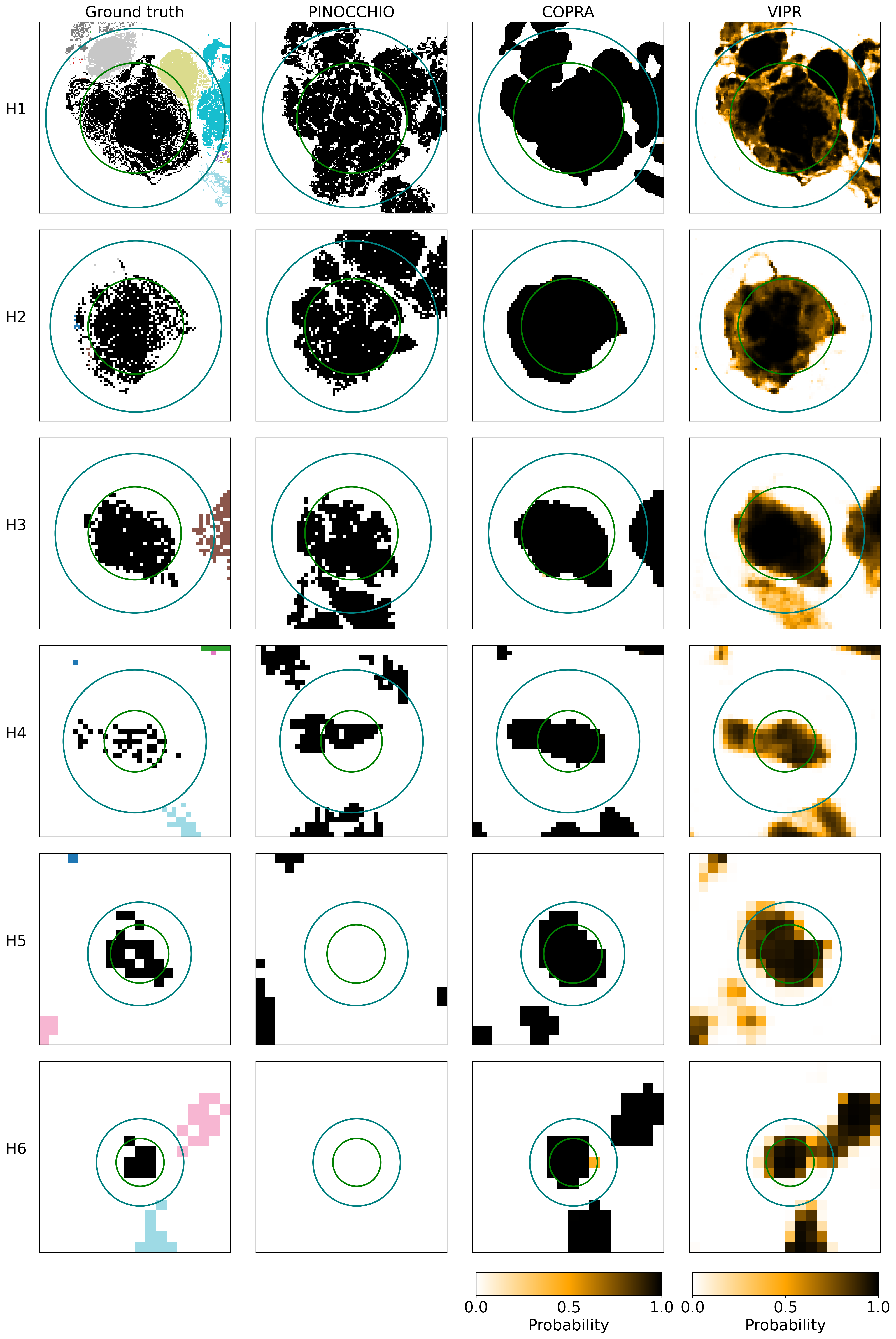}
    \caption{As in Figure~\ref{fig:visual_res}, but for the six representative halos described in Table~\ref{tab:objects_errors}.
    Each slice intersects the respective proto-halo patch along a plane containing the center of mass and oriented perpendicular to one of the Cartesian axes of the simulation box.
In each panel, the inner circle denotes the Lagrangian radius corresponding to a spherical proto-halo of that mass, while the outer circle indicates the maximum proto-halo radius $R_\mathrm{max}$, defined as the farthest distance from the center of mass to any voxel in the ground truth mask.
In the leftmost column showing the ground truth segmentation, voxels belonging to the target proto-halo are shown in black, while coloured voxels mark particles associated with other nearby halos in the same mass bin.}
    \label{fig:individual-halos}
\end{figure}

As in figure~\ref{fig:visual_res}, figure~\ref{fig:individual-halos} presents single-voxel slices for the proto-halos H1 to H6 comparing the ground truth segmentation (leftmost column) to the soft probabilistic predictions from the neural networks and the output of \textsc{pinocchio}. The differences in prediction quality are immediately apparent: \textsc{pinocchio} exhibits the poorest agreement with the ground truth, frequently generating false positives, introducing spurious features, and in some cases entirely missing objects --- most notably H5 and H6, which do not appear as any recognizable structures in its output. The figure also highlights differences in how models handle the complexity of proto-halo environments. Among the examples, H2 stands out as a particularly irregular object, characterized by numerous cavities rather than a single connected patch. This fragmented morphology proves difficult for the models to capture: \textsc{vipr} reflects the patchy membership through spatially varying probabilities, whereas \textsc{copra} and \textsc{pinocchio} fail to reproduce this intricate structure. Overall, \textsc{copra} successfully captures the general morphology of the proto-halos but lacks precision in finer structures. In contrast, \textsc{vipr} achieves excellent agreement with the ground truth, accurately reconstructing even irregular boundaries and detailed internal features.
\begin{table}
    \small
    \centering
    \begin{tabular}{c c c c c c c}
\hline
        Label & $\log\left(\frac{M}{h^{-1}\mathrm{M}_\odot}\right)$ & \# particles & $R_\mathrm{Lag}$ &
        $R_{\mathrm{max}}$ \\
        & & &$[h^{-1}\mathrm{Mpc}]$ & 
        $[h^{-1}\mathrm{Mpc}]$ \\
        \hline
        H1 & 14.38 & 369,051 & 8.71 & 14.09 \\
        H2 & 13.55 & 54,587 & 4.61 & 8.22 \\
        H3 & 12.79 & 9,583 & 2.57 & 4.47 \\
        H4 & 11.82 & 1,016 & 1.22 & 2.85 \\
        H5 & 10.89 & 119 &0.60 & 1.06 \\
        H6 & 10.49 & 48 & 0.44 & 0.80 \\
        \hline
    \end{tabular}
    \caption{Properties of the six representative halos selected for detailed object-level evaluation.
}
    \label{tab:objects_errors}
\end{table}

A more quantitative evaluation is provided in table~\ref{tab:halo_masses}, where
we assess prediction accuracy at the level of individual objects. Specifically, we measure the fraction of the Lagrangian volume (within both the $R_\mathrm{Lag}$ and $R_\mathrm{max}$) that is assigned to the true mass class by each model. Although this quantity does not correspond exactly to the halo mass, it serves as a practical proxy for object-level accuracy in the absence of a fragmentation step that would identify distinct halos within the segmented regions.\footnote{In particular, for consistency across methods, we do not use the internal halo fragmentation provided by \textsc{pinocchio} but treat its output as a segmentation mask in the same way as for the neural networks. }The interpretation of these fractions, however, requires some care: while the volume enclosed within $R_\mathrm{Lag}$ is typically well defined and only mildly influenced by neighboring structures, the larger region within $R_\mathrm{max}$ often receives significant external contributions from surrounding halos, which makes its correspondence to the true halo mass less direct.
For the three least massive classes, instead of presenting results for single objects, we report averages over all halos in the test simulation that have the same masses as H4, H5, and H6. These ensembles, denoted E4, E5, and E6, respectively, contain 14, 329, and 1594 halos. The averaging is performed over the individual volume fractions of each halo, accounting for their different values of $R_\mathrm{max}$, thus ensuring a statistically robust characterization of model performance in the low-mass regime.
The analysis of the predicted Lagrangian volume fractions, reported in table~\ref{tab:halo_masses}, reveals several clear trends. The true fraction of voxels belonging to the proto-halos within $R_\mathrm{Lag}$ is about 0.75 for the three most massive halos (H1–H3) and then decreases rapidly with mass, reaching roughly 0.38 for the smallest halos. This behavior reflects that low-mass proto-halos are less spherical and display increasingly irregular, “frothy” or “spongy” structures with internal voids. When the same fraction is computed within $R_\mathrm{max}$, its dependence on mass is more erratic, likely because the position of the most distant voxel from the halo center is sensitive to random fluctuations. Nevertheless, the values for the ground truth remain between 0.15 and 0.4 for all halo masses.
Both \textsc{copra} and \textsc{vipr} reproduce these trends well, but they systematically overestimate the volume fractions --- and therefore the halo masses --- typically by about 10\% and occasionally up to 20\%. This overestimation appears consistently in the measurements at both $R_\mathrm{Lag}$ and $R_\mathrm{max}$.
The results obtained with \textsc{pinocchio} are less regular. For the smallest halos and within $R_\mathrm{Lag}$, its accuracy is significantly worse than that of the neural networks, highlighting its limitations in capturing the complex morphology of low-mass proto-halos. In contrast, when evaluated at $R_\mathrm{max}$, the performance of \textsc{pinocchio} approaches that of the neural networks, although it still shows larger scatter.
In conclusion, the neural networks consistently outperform \textsc{pinocchio}, particularly for low-mass halos and within $R_\mathrm{Lag}$, while preserving the qualitative behavior observed in the ground truth distributions.

\begin{table}
    \small
    \centering
    \begin{tabular}{c c c c c c c c c c}
\hline
  & \multicolumn{4}{c}{$f_V(<R_\mathrm{Lag})$}& &\multicolumn{4}{c}{$f_V(<R_\mathrm{max})$}\\
Label & Ground & \textsc{pinoc.} & \textsc{copra} & \textsc{vipr} & &
Ground & \textsc{pinoc.} & \textsc{copra} & \textsc{vipr}\\
& truth & & & & & truth & & & \\
        \hline
        H1 & 0.750 & 0.786 & 0.882 & 0.834 & & 0.356 & 0.454 & 0.429 & 0.404 \\
        H2 & 0.719 & 0.901 & 0.850 & 0.769 & & 0.177 & 0.290 & 0.223 & 0.211 \\
        H3 & 0.786 & 0.683 & 0.853 & 0.827 & &0.228 & 0.245 & 0.236 & 0.226 \\
        \hline
        E4 & 0.578 & 0.719 & 0.675 & 0.623 & &0.154 &0.198  &0.179 & 0.163 \\
        E5 & 0.486 & 0.247 & 0.642 & 0.558 & &0.399 &0.458  &0.415 & 0.489 \\
        E6 & 0.376 & 0.501 & 0.459 & 0.355 & &0.295 &0.368  &0.334 & 0.323 \\
        \hline
    \end{tabular}
    \caption{Fraction of the Lagrangian volume within $R_\mathrm{Lag}$ and $R_\mathrm{max}$ (measured from the true proto-halo center) that
    is assigned to the correct mass class by each model.
    }
    \label{tab:halo_masses}
\end{table}

\begin{figure}
    \centering
    \includegraphics[width=0.7\linewidth]{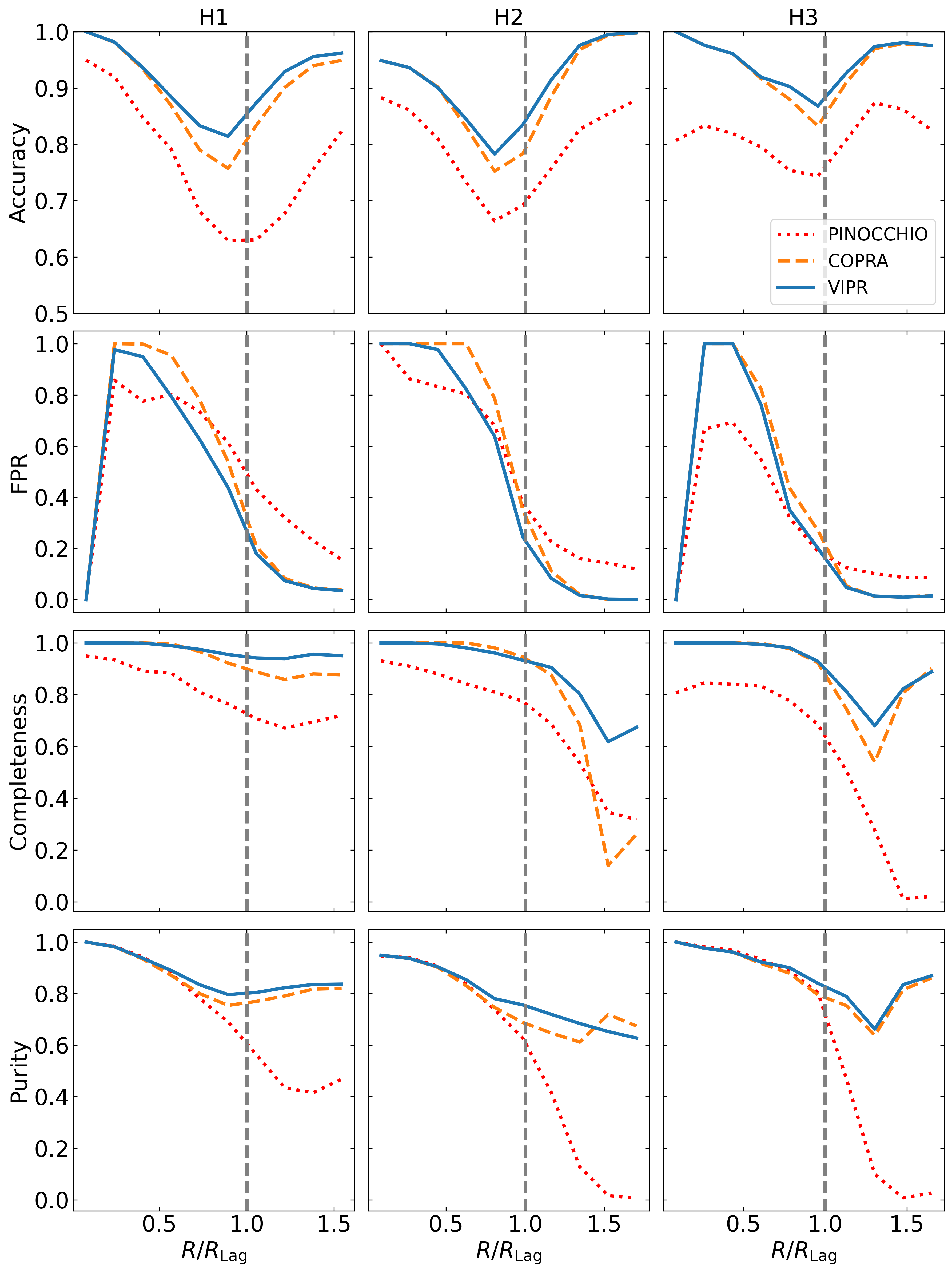}
    \caption{Radial performance profiles for three representative proto-halos (H1, H2, H3 from table~\ref{tab:objects_errors}; columns from left to right). Rows display (top to bottom) accuracy, false positive rate (FPR), completeness, and purity as functions of distance from the true center of mass. Solid, dashed, and dotted lines correspond to \textsc{vipr}, \textsc{copra}, and \textsc{pinocchio}, respectively. The vertical dotted line marks the Lagrangian radius $R_\mathrm{Lag}$. The profiles are shown out to the maximum proto-halo extent $R_\mathrm{max}$.  }
    \label{fig:radius_plots_single}
\end{figure}

Figure~\ref{fig:radius_plots_single} presents spherically averaged radial profiles for three representative proto-halos (H1, H2, H3; left to right), illustrating how model performance varies with distance from the true center of mass. The rows correspond to four metrics: accuracy, FPR, completeness, and purity (top to bottom). Different linestyles denote the models: solid for \textsc{vipr}, dashed for \textsc{copra}, and dotted for \textsc{pinocchio}. A vertical dashed line marks the Lagrangian radius $R_\mathrm{Lag}$ of each halo.
Across all halos, accuracy is highest near the centers, decreases approaching $R_\mathrm{Lag}$, and rises again in the outer background where classification is less ambiguous. \textsc{vipr} consistently delivers the highest accuracy, with \textsc{copra} close behind, while \textsc{pinocchio} performs notably worse.
The FPR peaks at small radii (where the metric is sensitive to a few misclassified voxels) and drops rapidly outward. Beyond $R_\mathrm{Lag}$, \textsc{pinocchio} exhibits substantially elevated FPR compared to the neural networks, reflecting its tendency to overpredict halo membership in the background.
For completeness, both neural networks achieve values close to unity within $R_\mathrm{Lag}$, demonstrating their ability to recover almost all halo material. Outside this radius, completeness declines moderately, with \textsc{vipr} maintaining slightly higher values than \textsc{copra}. In contrast, \textsc{pinocchio} shows consistently lower completeness at all radii.
Finally, purity starts near one at the center but decreases toward the outskirts due to contamination from false positives. The neural networks maintain moderate purity (around 0.7–0.8) in the outer regions, while \textsc{pinocchio} --- although comparable within $R_\mathrm{Lag}$ --- suffers a sharp purity drop beyond this boundary.
These results confirm that the deep learning models, particularly \textsc{vipr}, achieve superior performance across all metrics and radii compared to the perturbation-theory-based \textsc{pinocchio}.

\begin{figure}
    \centering
    \includegraphics[width=0.7\linewidth]{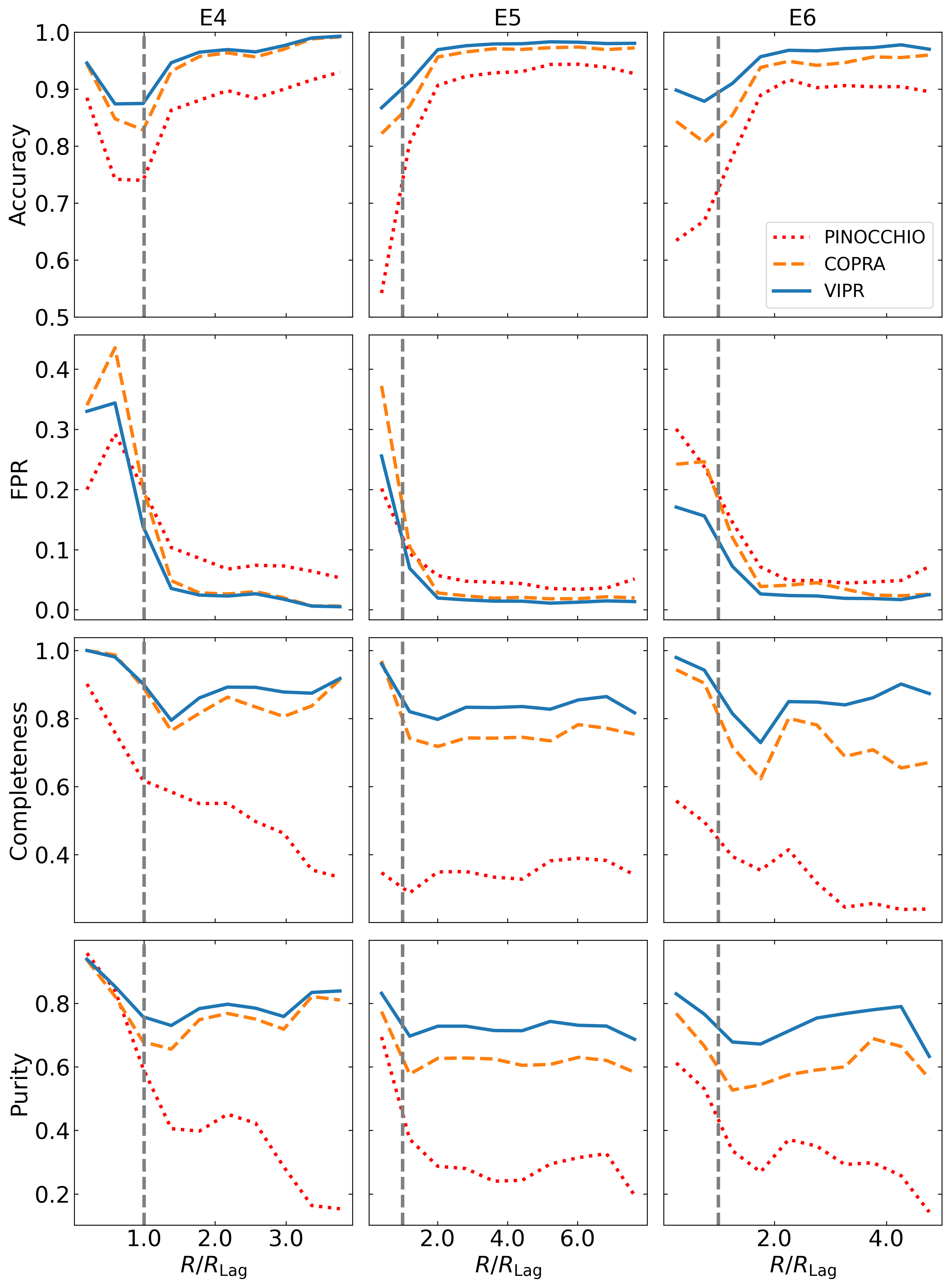}
    \caption{Same as in figure~\ref{fig:radius_plots_single}, but for
    the small ensembles of galaxy-sized halos with identical mass
    E4, E5, and E6.}
    \label{fig:radius_plots_many}
\end{figure}

Figure~\ref{fig:radius_plots_many} presents the same radial analysis as in the previous figure, but now focuses on smaller halos. For these cases, individual profiles are too noisy to be informative due to the small size of the objects in Lagrangian space. To improve statistical robustness, we average the profiles over the ensembles E4, E5, and E6. For the neural networks, the trends mirror those observed for higher-mass halos: \textsc{vipr} consistently delivers the best performance, closely followed by \textsc{copra}. Both models maintain high accuracy and completeness within $R_\mathrm{Lag}$ and experience only moderate declines beyond it, with \textsc{vipr} showing the most stable behavior. In stark contrast, \textsc{pinocchio} performs poorly across the entire radial range for these low-mass objects. Its completeness is significantly reduced even well inside $R_\mathrm{Lag}$, and its purity is already degraded at small radii before dropping steeply outside the halo boundaries. These results emphasize the increasing difficulty of the perturbation-theory-based model in correctly identifying and classifying low-mass proto-halos.

\subsection{Training with different inputs}

In some cases, modifying the input features of a deep neural network can reveal how different physical quantities contribute to the model's predictive performance. For instance, analyzing the effect of derivatives or transformations of the density field can provide insights into which aspects of the input are most informative for predicting halo formation.

Although certain fields --- such as the gravitational potential --- are formally related to the density field via a convolution, our models operate on small input volumes that represent only a tiny fraction of the full simulation domain. Specifically, the gravitational potential $\Phi(\mathbf{x})$ satisfies the Poisson equation
$\nabla^2 \Phi = \delta$,
whose solution involves convolving the density contrast $\delta$ with the Newtonian Green’s function $G(\mathbf{x}) = -1/|\mathbf{x}|$. This defines a non-local relation that depends on long-range correlations across the entire volume. However, individual samples from our dataset are small enough to violate the periodic boundary conditions of the full box and cannot reliably capture these global dependencies. As a result, the network has no access to this large-scale information when trained solely on localized density patches. Incorporating derived fields such as the gravitational potential or tidal shear as additional inputs may therefore provide valuable complementary information and enhance the model’s predictive capacity.

To investigate whether this approach can offer physical insight into halo formation, we experiment with including a scalar representation of the tidal shear field. Specifically, we compute the traceless part of the tidal tensor (the tidal shear) 
\begin{equation}
    T_{ij}=\frac{\partial^2 \Phi}{\partial x_i \partial x_j}-\frac{1}{3}\nabla^2 \Phi \,\delta_{ij}\;,
    \end{equation}
    (where $\delta_{ij}$ denotes the Kronecker symbol)
and derive a scalar input by taking its Frobenius norm,
\begin{equation}
 T=\sqrt{\sum_{i=1}^3 \sum_{j=1}^3 (T_{ij})^2}\;.
 \label{eq:frobenius}
\end{equation}
This scalar field captures the anisotropic component of local gravitational interactions, quantifying how matter collapses or stretches along preferred directions, while 
excluding isotropic compression or dilation (i.e.,  uniform expansion or contraction encoded in the matter overdensity). 
Importantly, the resulting field 
$T$ is non-Gaussian even when derived from an underlying Gaussian density field; its statistical properties, including its probability distribution, are discussed in detail in appendix~\ref{appendix:frob}.

We train both \textsc{copra} and \textsc{vipr} under two input configurations: first, using only the tidal shear field, and second, combining the tidal shear with the density field. Validation accuracies and AUC scores for these experiments are reported in Tables \ref{tab:diff_inputs_val} and \ref{tab:AUC_tides}, allowing a direct comparison of model performance across input combinations. The consistently high AUC values demonstrate that the tidal shear alone encodes substantial predictive information. For \textsc{copra}, a modest increase in AUC is observed when both tidal shear and density are provided, suggesting that the two fields may contribute weakly complementary information, even though the effect remains small.

\begin{table}
    \centering
    \begin{tabular}{c c c}
    \hline
        Input & \textsc{copra}& \textsc{vipr} \\
        \hline
         $\delta$
         & 85.2& 92.4\\
         $T$ & 83.7 & 92.3\\
         $\delta \land T$
         & 85.8& 93.1\\
         \hline
    \end{tabular}
    \caption{Validation accuracies (in percent) of models with seven output classes, trained on different input configurations. Including both the density field and tidal shear as inputs leads to improved accuracy, despite the models having identical learning capacity (i.e., same architectural complexity and number of trainable parameters) across all configurations.}
    \label{tab:diff_inputs_val}
\end{table}

\begin{table}[]
    \centering
    \begin{tabular}{cccccccccc}
    \hline
    Model & Input & \multicolumn{6}{c}{Halo mass bin} & Binary mask\\
           &  & 1 & 2 & 3 & 4 & 5 & 6 &\\
             \hline
     & $\delta$ &0.93 & 0.94 & 0.94 & 0.95 & 0.95 & 0.95 & 0.95\\   
     \textsc{copra} & $T$ & 0.92 & 0.93 & 0.93 & 0.93 & 0.93 & 0.93 & 0.93\\ 
     & $\delta \land  T$ & 0.94 & 0.95 & 0.95 & 0.95 & 0.95 & 0.95 & 0.95\\ 
    \hline
    & $\delta$ & 0.99 & 0.99 & 0.99 & 0.99 & 0.99 & 0.99 & 0.99\\   
     \textsc{vipr} & $T$ & 0.99 & 0.99 & 0.99 & 0.99 & 0.99 & 0.99 & 0.99\\ 
     & $\delta\land T$ & 0.99 & 0.99 & 0.99 & 0.99 & 0.99 & 0.99 & 0.99\\ 
    \hline
    \end{tabular}
    \caption{AUC scores for the 7 output classes of \textsc{copra} and \textsc{vipr}, trained on dataset 3. Results are presented for three different input configurations: the density field alone, the tidal shear alone, and a combination of both.}
    \label{tab:AUC_tides}
\end{table}

\subsection{Inspecting class-activation maps with Grad-CAM}

To further explore the interpretability of our neural network models, we implement the Grad-CAM algorithm \cite{gradcam} to visualize class-activation maps for \textsc{copra}, our CNN-based architecture. Class-activation maps represent the output of the final convolutional layer---producing one map per class --- and highlight the spatial regions of the input that contribute most strongly to a given prediction. Grad-CAM enhances these maps by incorporating gradient information, which quantifies how changes in the input affect the output for a specific class. The resulting heatmaps reveal which input features are most influential in the model’s decision-making process.

\begin{figure}
    \centering    \includegraphics[width=0.8\linewidth]{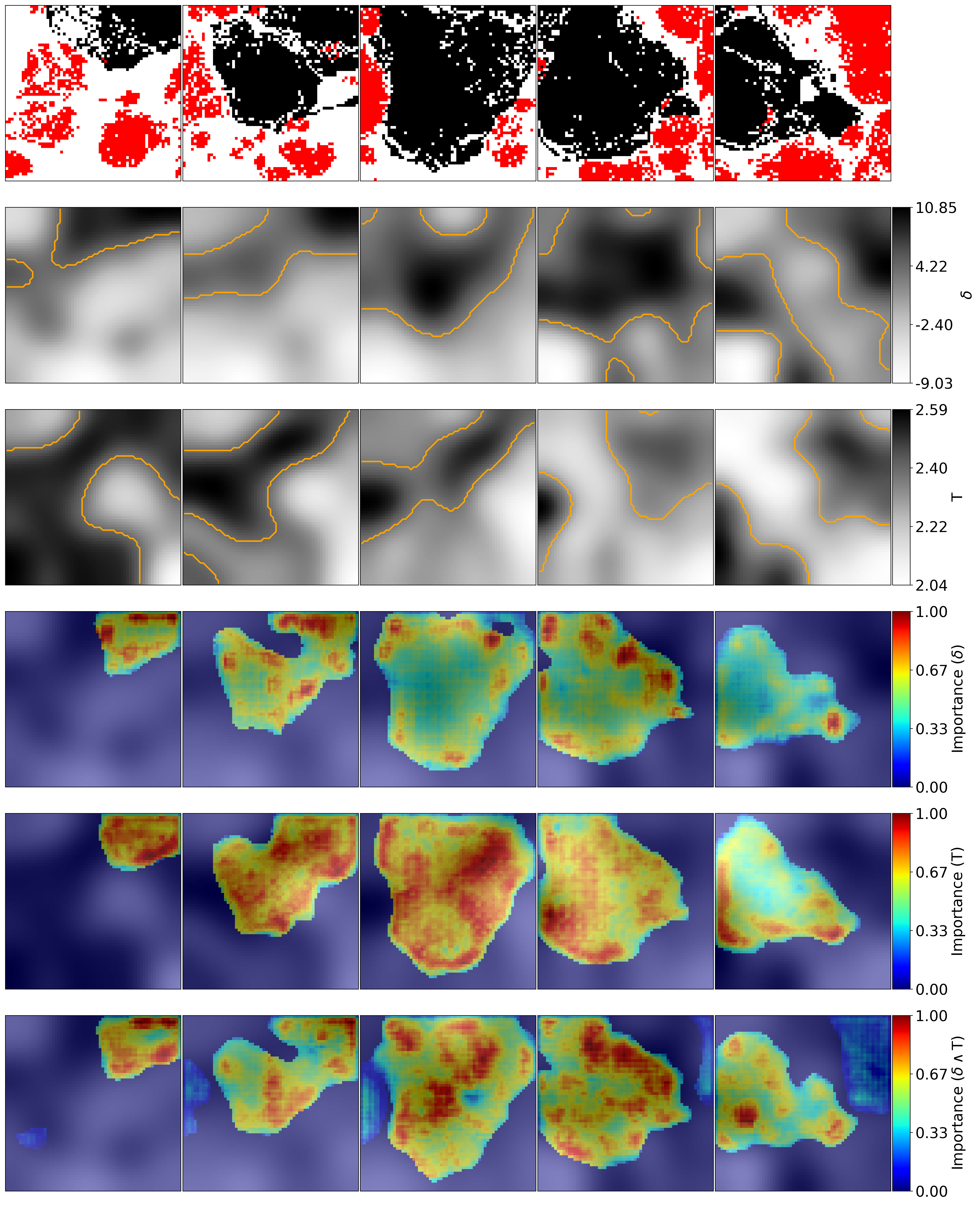}
    \caption{Grad-CAM analysis of a high-mass proto-halo from the test simulation.
    Each column shows a 2D slice intersecting the 3D proto-halo patch at planes spaced 12 voxels apart along one Cartesian axis. The top row displays the ground truth segmentation; colored voxels indicate halo membership, with black representing the target halo and red marking neighboring structures. The second and third rows show the smoothed input fields: the overdensity $\delta$ and the Frobenius norm of the tidal shear $T$, respectively, both linearly extrapolated to $z=0$ and smoothed with a Gaussian filter of $5.48\,h^{-1}$ Mpc.
    The orange contours correspond to $\delta=1.686$ and $T$ one standard deviation above the mean   
    (see appendix~\ref{appendix:frob}).
    The bottom three rows present Grad-CAM heatmaps for models trained on $\delta$, $T$, and both combined. Each heatmap indicates the relative importance of different regions in driving the network’s class prediction.  }
    \label{fig:cam_3_big}
\end{figure}

In figure~\ref{fig:cam_3_big}, we explore the Lagrangian region of a massive halo ($M=1.32\times 10^{14}\,h^{-1}$ M$_\odot$) to identify consistent patterns that the network relies on to infer specific halo mass classes. To visualize model behavior in a three-dimensional volume, we present a tomographic sequence of two-dimensional slices, evenly spaced along the third dimension. In this case, five slices are shown, each separated by 12 voxels, providing a representative scan through the proto-halo patch.
The top row displays the ground truth segmentation from the test simulation (black is the proto-halo patch, red are voxels in other proto-halos, white are non-halo voxels). The second and third rows show the two possible input fields for \textsc{copra}: the overdensity field ($\delta$) and the scalar tidal-shear field ($T$), both linearly extrapolated to $z=0$ and smoothed with a Gaussian filter over a scale of $R_\mathrm{G}=5.48\,h^{-1}$ Mpc.
The smoothing scale is chosen so that the mass enclosed by the filter, $M=(2\pi)^{3/2}\,\bar{\rho}\,R_\mathrm{G}^3$, matches the halo mass. For reference, contours corresponding to $\delta=1.686$ (the critical overdensity in a Press-Schechter-like approach) and to $T$ one standard deviation above the mean (see appendix~\ref{appendix:frob}) are overlaid to indicate typical regions that would be expected to collapse under simplified models or with enhanced tidal strength.
The bottom three rows present the Grad-CAM heatmaps generated from CNNs trained on different input configurations: the density field alone, the tidal shear alone, and both $\delta$ and $T$ together. These heatmaps highlight the spatial regions that most strongly influence the model's prediction for the selected mass class. By comparing activation patterns across different input modalities, we gain insight into the distinct features each input contributes to the model's decision-making process.
To aid comparison, the heatmaps are overlaid on grayscale maps of the input fields ($\delta$ for the top and bottom heatmaps, $T$ for the middle one).

The figure reveals distinct spatial patterns in the Grad-CAM activations depending on the input used.
When using only the density field as input, the heatmaps show 
numerous localized peaks concentrated near proto-halo boundaries, sometimes aligned with local density extrema (maxima and minima).
This suggests that the model trained on density relies heavily
on edge-like features to delineate halo regions. In contrast, the heatmaps derived from the tidal shear input 
exhibit diffuse, interconnected hot regions that often permeate proto-halo interiors and only occasionally emphasize boundaries. 
This pattern indicates that the tidal shear input leads the network to exploit information embedded in the internal structure of the proto-halos rather than relying predominantly on their edges. For the model trained on both density and tidal shear, the resulting heatmaps show an intermediate behavior, combining boundary sensitivity with some degree of interior activation. Despite these differences in the highlighted regions, all three examined proto-halo patches reproduce the ground truth segmentation quite faithfully, reflecting the overall robustness of the models.

\begin{figure}
    \centering
    \includegraphics[width=0.8\linewidth]{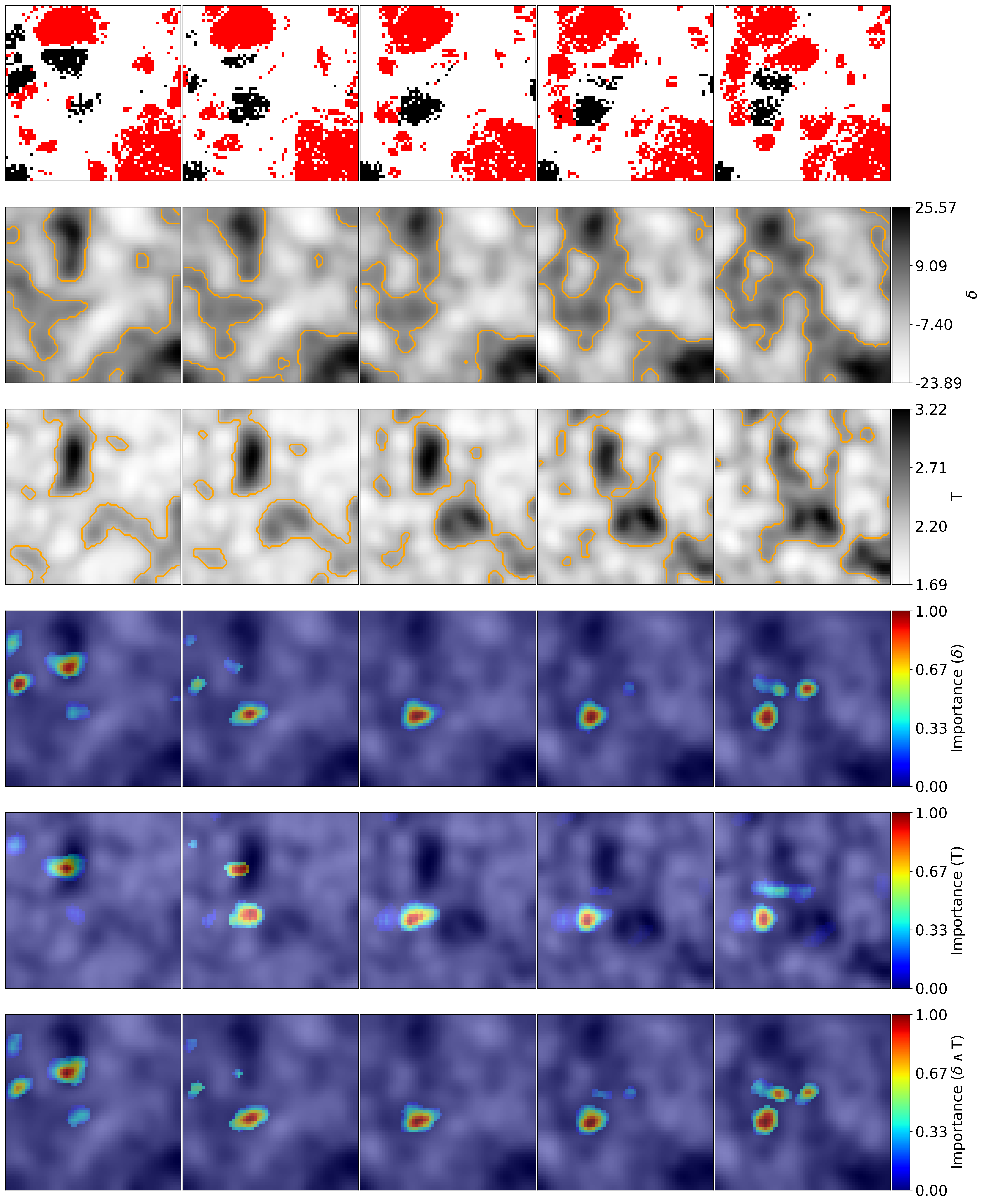}
    \caption{Same as figure~\ref{fig:cam_3_big}, but for a galaxy-sized proto-halo. Owing to the compact size of the object, the slices are spaced only 2 voxels apart, and the input fields are smoothed with a Gaussian filter of $R_\mathrm{G}=0.836\,h^{-1}$ Mpc.}
    \label{fig:cam_3_small}
\end{figure}

In figure~\ref{fig:cam_3_small}, we extend the Grad-CAM analysis to a galaxy-sized proto-halo of mass $M=5.62\times 10^{11}\,h^{-1}$ M$_\odot$. Because of its compactness,
we present five slices each separated by 2 voxels. 
The proto-halo appears as the central cluster of black pixels in the images, while other black regions correspond to additional objects belonging to the same mass class.
The input fields are smoothed using a Gaussian kernel with a scale of $R_\mathrm{G}=0.836 \,h^{-1}$ Mpc.
The resulting heatmaps show a sharp, compact activation 
at the proto-halo location. 
The model trained on the density field shows enhanced gradients in regions of intermediate extended overdensity. When trained on tidal shear, the strongest responses occur where $T$ has a local minimum 
flanked by peaks. Using both $\delta$ and $T$ 
yields activation patterns broadly similar to the other cases.

Interpreting these patterns remains challenging. While the activation maps suggest some spatial correlation between the network’s attention and physical features of the input fields, firm conclusions cannot be drawn from visual inspection alone. 
To obtain more quantitative insights, we cross-correlated the heatmaps with the input fields (and the magnitudes of their gradients) smoothed at various scales. Although correlations peak at small smoothing scales, they remain weak, with Spearman coefficients around 0.2–0.3. This suggests that extracting clear physical interpretation from Grad-CAM maps is far from trivial.

\section{Summary \& Conclusions}
\label{sec:summary}

In this work, we trained and evaluated two deep learning models for segmenting protohalos from initial conditions of N-body simulations and classifying them into discrete halo mass bins.
The first, \textsc{copra}, employs a CNN-based V-net architecture, while the second, \textsc{vipr}, uses a ViT-based UNETR architecture. 
Both models are trained using the density field as input.
Our findings clearly demonstrate the superior performance of the transformer-based \textsc{vipr}, which consistently achieves validation accuracies exceeding 90\% and AUC scores above 0.99 across all output classes. These results highlight the promise of vision transformers for precision structure formation modeling, outperforming more conventional convolutional approaches.

On the scale of the entire test simulation box, both neural networks provide highly accurate reconstructions of the total halo mass per class, with sub-percent deviations from the ground truth, compared to significantly larger errors from the perturbation-theory-based method \textsc{pinocchio} (ranging from 10\% to 60\% depending on the mass bin). 
Importantly, this performance persists at the level of individual objects. \textsc{vipr} yields mass estimates within 10\% of the true value even for challenging, irregular protohalo shapes. Visual inspections and voxel-wise radial profiles confirm that \textsc{vipr} accurately reconstructs detailed internal structure and boundaries, while \textsc{copra} provides a robust but less refined segmentation. 
Because these models are fully convolutional and inherently scalable, they can in principle be applied to simulation boxes of any size, provided the resolution remains unchanged. In practice, this would require expanding the training set to include halos covering the wider mass range present in larger volumes, but the overall prediction quality is not expected to degrade significantly.

Our study employs relatively coarse mass binning, since the combination of small sub-volumes and a limited number of simulation boxes would otherwise leave finer bins sparsely populated. We therefore frame the task as a softmax classification problem, while noting that the same architectures could be extended to regression with deeper networks and larger datasets. In this sense, our work should be viewed as a proof of concept, demonstrating the potential of vision transformers for protohalo segmentation and paving the way for future large-scale regression-based studies.

We also experimented with alternative input configurations.
Specifically, we trained models using only the Frobenius norm of the tidal shear field, a scalar metric for anisotropic gravitational effects.
We found that tidal shear alone carries significant predictive information.
When both the density field and the tidal shear were used as input, models achieved better performance despite unchanged learning capacity, indicating complementary, non-redundant contributions of each field. This suggests that including physically motivated fields can enhance model performance without overfitting.

To further probe model behavior, we employed Grad-CAM to generate class-activation maps from \textsc{copra}. For the most massive halos, these heatmaps reveal distinct patterns depending on the input: training on the density field produces localized activations concentrated around proto-halo boundaries, whereas training on the tidal shear yields more diffuse responses that percolate through the interior of the proto-halos.  Although we attempted to quantify these patterns by cross-correlating the heatmaps with the input fields and their gradients at various smoothing scales, the correlations remained weak. This suggests that the information captured by the network is not trivially linked to simple features of the input fields. While the insights from Grad-CAM remain qualitative, they still provide a useful starting point for exploring how different inputs influence the network’s decision-making.

The primary limitation of our approach is that 
we do not address the task of fragmenting the predicted segmentation maps into individual halos. This post-processing step is essential for direct object-level comparison to halo catalogs and constitutes a non-trivial challenge.

Finally, we emphasize that the primary goal of this work is to assess the viability of deep neural networks as fast, high-accuracy surrogates for structure formation modeling. While interpretability remains limited due to model complexity, this trade-off is acceptable in contexts where precision and speed are paramount. In future work, we are interested in integrating approaches such as deep symbolic regression \cite{symbolicregression}, which may offer a path toward more interpretable yet expressive models capable of uncovering analytic structure–formation relationships.

\acknowledgments
This work includes research conducted as part of the Master’s thesis of TA, completed in 2023 at the University of Bonn. The majority of this work was completed using the Bender GPU cluster at the University of Bonn. The development of \textsc{copra} was implemented entirely in \textsc{TensorFlow} using the \textsc{Keras} backend. We gratefully acknowledge the developers of \textsc{self-attention-cv}, which provided the foundation for implementing the \textsc{vipr} model. We also thank the authors of the simulation and analysis codes \textsc{Gadget}, Amiga Halo Finder (AHF), and \textsc{pinocchio} for making their software publicly available and well-documented. We are grateful to Zorah Lähner for her constructive feedback on an earlier version of this manuscript. TA is a member of the International Max Planck Research School (IMPRS) for Astronomy and Astrophysics and acknowledges financial support from the Bonn-Cologne Graduate School of Physics and Astronomy (BCGS).
\\ \\
\textbf{Code \& data availability}\\ \\
The implementations of both neural network architectures, along with the scripts used to generate training data and reconstruct the test simulation cubes from model predictions, are publicly available at: 
\github{https://github.com/tokaalokda/DProtohalos} \href{https://github.com/tokaalokda/DProtohalos}{github.com/tokaalokda/Dprotohalos}.
The specific datasets and the best-performing model checkpoints used in this study are not publicly hosted but can be made available by the authors upon reasonable request.

\appendix
\section{Voxel-wise agreement}
\label{appendix:voxel_agreement}
\begin{figure}
    \centering
    \includegraphics[scale=0.24]{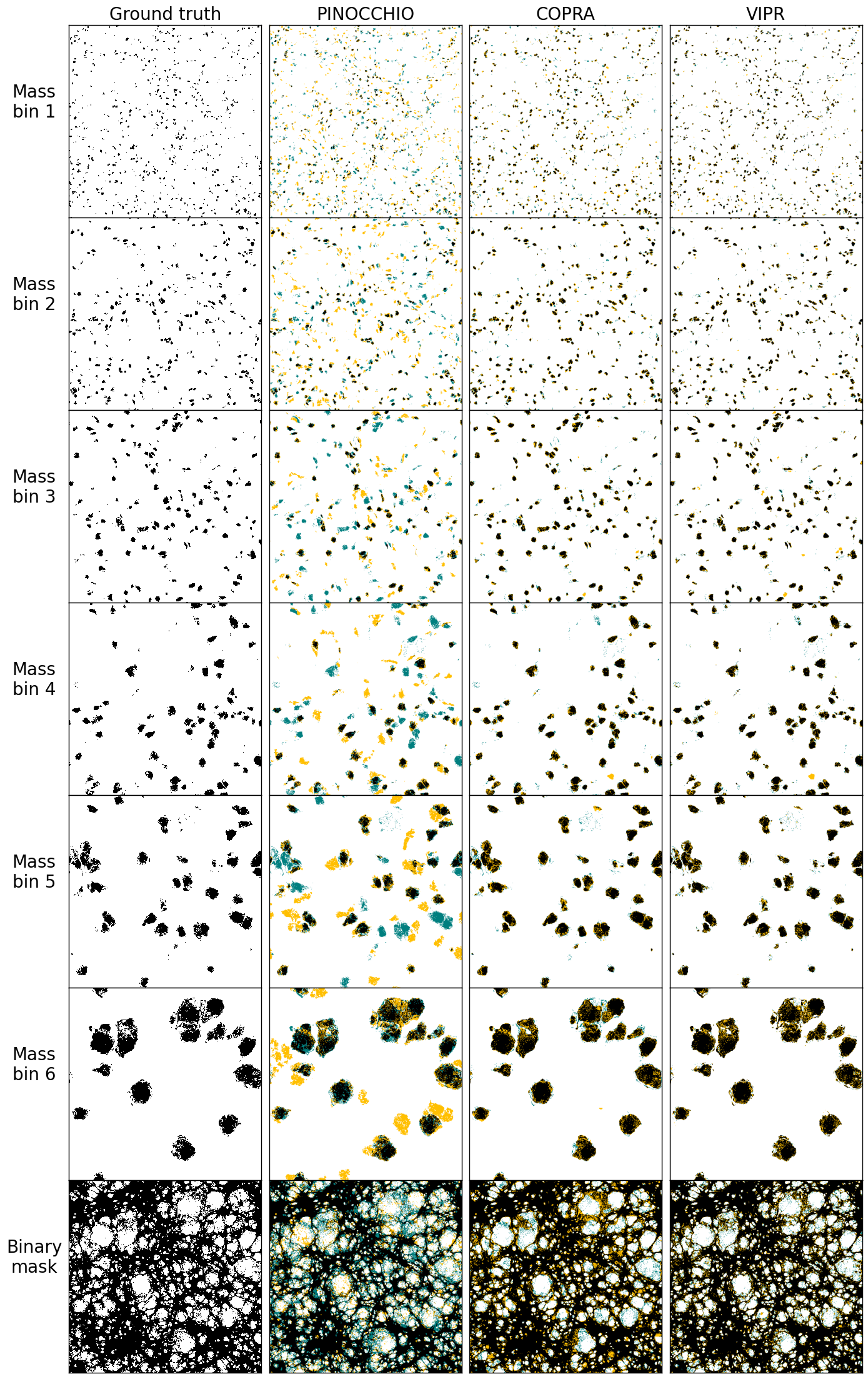}
    \caption[Short caption]{
    Same visualization setup as in figure~\ref{fig:visual_res}, but using hard classifications to explicitly distinguish between prediction outcomes. Voxels are colored according to their classification result: true positives (black), true negatives (white), false positives (amber), and false negatives (teal). This representation highlights the spatial distribution of correct and incorrect predictions across the segmented region.}
    \label{fig:visual_res_color}
\end{figure}

To further illustrate the classification performance of our models, we provide in figure~\ref{fig:visual_res_color} a voxel-wise visualization of prediction outcomes using hard classifications. This figure complements the soft-probability maps shown earlier by explicitly distinguishing true positives, true negatives, false positives, and false negatives. By doing so, it highlights the spatial distribution of correct and incorrect predictions within and around the proto-halo regions, offering additional insight into typical failure modes and boundary uncertainties.

Notably, the neural networks yield significantly more accurate and spatially coherent classifications than the \textsc{pinocchio} model. In contrast, \textsc{pinocchio} produces a considerably higher number of both false positives and false negatives --- reflecting its tendency to simultaneously overpredict the total collapsed mass and misclassify substantial portions of the true proto-halo structures. In many cases, false positives in a given mass class are mirrored by false negatives in neighboring bins, consistent with \textsc{pinocchio}’s systematic overestimation of individual halo masses.

\section{Frobenius norm of the linear tidal shear}
\label{appendix:frob}
We assume that the linear mass overdensity field $\delta(\mathbf{x})$ forms
a statistically homogeneous and isotropic Gaussian random field with zero mean and
power spectrum $P(k)$ defined by
\begin{equation}
    \langle \tilde{\delta}(\mathbf{k})\,\tilde{\delta}(\mathbf{k'})\rangle
    =(2\pi)^3\,\delta^{(3)}_\mathrm{D}(\mathbf{k}+\mathbf{k}')\,P(k)\;,
\end{equation}
where
\begin{equation}
    \tilde{\delta}(\mathbf{k})= \int \delta(\mathbf{x})\,e^{i\mathbf{k}\cdot\mathbf{x}}\,\mathrm{d}^3x
\end{equation}
is the Fourier transform of $\delta(\mathbf{x})$ and
$\delta^{(3)}_\mathrm{D}(\mathbf{k})$ denotes the Dirac delta function in three dimensions. 
After applying a spherically symmetric low-pass filter with window function $\widetilde{W}(kR)$,  
the variance of the smoothed field becomes
\begin{equation}
    \sigma^2_R\equiv\langle\delta_R^2(\mathbf{x}) \rangle=\frac{1}{2\pi^2}\,\int_0^\infty k^2\,P(k)\,\widetilde{W}^2(kR)\,\mathrm{d}k\;.
\end{equation}

\subsection{Variance and covariance of the shear components}

The tidal shear tensor is defined as the traceless part of the Hessian of the potential. In Fourier space, its components are
\begin{equation}
    \widetilde{T}_{ij}(\mathbf{k})=   
    \left(\hat{k}_i\,\hat{k}_j-\frac{1}{3}\,\delta_{ij} \right)\,\tilde{\delta}_R(\mathbf{k})\;,
\end{equation}
where $\hat{k}_i=k_i/k$.
The variance of $T_{ij}$ is
\begin{align}
    \langle T_{ij}^2(\mathbf{x})\rangle&=
    \int \left(\hat{k}_i\hat{k}_j-\frac{1}{3}\,\delta_{ij}\right)^2P(k)\,W^2(kR)\, \frac{\mathrm{d}^3k}{(2\pi)^3}\nonumber\\
    &=A_{ij}\,\sigma^2_R\;,
\end{align}
where 
\begin{equation}
    A_{ij}=\int_{4\pi}  \left(\hat{k}_i\hat{k}_j-\frac{1}{3}\,\delta_{ij}\right)^2\,\frac{\mathrm{d}^2\hat{k}}{4\pi}=
    \begin{cases}
        \frac{4}{45} & \text{if $i=j$} \\
        \frac{1}{15}& \text{otherwise}
    \end{cases}
\end{equation}
is a geometric factor.
For distinct diagonal terms ($i\neq j$), the covariance is
\begin{align}
    \langle T_{ii}(\mathbf{x})\,T_{jj}(\mathbf{x})\rangle&=
    \int \left(\hat{k}_i\hat{k}_i-\frac{1}{3}\right)\,\left(\hat{k}_j\hat{k}_j-\frac{1}{3}\right)P(k)\,W^2(kR)\, \frac{\mathrm{d}^3k}{(2\pi)^3}\nonumber\\
    &=\left(\frac{1}{15}-\frac{2}{3}\cdot\frac{1}{3}+\frac{1}{9} \right)\,\sigma^2_R=-\frac{2}{45}\,\sigma^2_R\;,
\end{align}
while non-matching off-diagonal terms are uncorrelated.

\subsection{Variance of the Frobenius norm}

The tidal shear tensor is trace-free, so
so its Frobenius norm squared
has expectation
\begin{align}
    \langle T^2(\mathbf{x})\rangle&=\langle T_{xx}^2(\mathbf{x})+T_{yy}^2(\mathbf{x})+[-T_{xx}(\mathbf{x})-T_{yy}(\mathbf{x})]^2+2[T_{xy}^2(\mathbf{x})+T_{xz}^2(\mathbf{x})+T_{yz}^2(\mathbf{x})]\rangle \nonumber \\
    &=2[\langle T_{xx}^2(\mathbf{x}) \rangle+\langle T_{yy}^2(\mathbf{x}) \rangle+\langle T_{xx}(\mathbf{x}) T_{yy}(\mathbf{x})\rangle+\langle T_{xy}^2(\mathbf{x})\rangle+\langle T_{xz}^2(\mathbf{x})\rangle+\langle T_{yz}^2(\mathbf{x})]\rangle] \nonumber \\
    &=\frac{2}{3}\,\sigma^2_R\;.
\end{align}

\subsection{Probability distribution of $T$}

The five independent components of the traceless tensor $T_{ij}$ that contribute to $T$ can
be linearly transformed into an orthonormal basis in which the quadratic form in eq.~(\ref{eq:frobenius}) becomes the sum of squares of five independent standard Gaussian variables. Consequently, $T$ follows a 
$\chi$ distribution with $k=5$ degrees of freedom,
\begin{equation}
    \mathcal{P}(T)\propto T^{k-1}\,\exp\left(-\frac{T^2}{2\sigma^2_\mathrm{c}}\right)\,
\end{equation}
where $\sigma^2_\mathrm{c}=\langle T^2\rangle/k$ is the variance of each of the five independent Gaussian variables that contribute to the Frobenius norm.
The corresponding mean, mode, and standard deviation of $T$ are
\begin{align}
    \langle T \rangle &=\frac{8\sqrt{2}}{3\sqrt{5\pi}}\,\sqrt{\langle T^2 \rangle}\simeq 0.672\,\sqrt{\langle T^2 \rangle}\approx 0.549\, \sigma_R \;,\\
    T_\mathrm{mode}&=2\,\sigma_\mathrm{c}=\frac{2}{\sqrt{5}}\,\sqrt{\langle T^2 \rangle}\approx 0.894\,\sqrt{\langle T^2 \rangle}\approx 0.730\, \sigma_R \;,\\
    \sigma_T&=\sqrt{\left(1- \frac{8\sqrt{2}}{3\sqrt{5\pi}}\right) \,\langle T^2 \rangle}\approx 0.741\,\sqrt{\langle T^2 \rangle}\approx 0.605 \,\sigma_R\;.
\end{align}

\subsection{Statistical relationship between $T$ and $\delta$ }
The components $T_{ij}$ of the linear tidal shear tensor are 
linearly related to $\delta$ and uncorrelated with it.
Since $\delta$ is a Gaussian random field, linear uncorrelation implies statistical independence, so $\delta$ 
and $T_{ij}$ are independent.
The Frobenius norm $T$, however, is a non-linear function of $T_{ij}$.
Although
$\langle \delta(\mathbf{x}) T^2_{ij}(\mathbf{x})\rangle=0$ (because this expectation involves a third-order moment of a zero-mean Gaussian random field) and therefore
$\langle T(\mathbf{x})\,\delta(\mathbf{x})\rangle=0$,
the conditional probability
$\mathcal{P}(T|\delta)$ still differs from the marginal $\mathcal{P}(T)$.
Thus, $\delta$ and $T$ are uncorrelated but not statistically independent.

\FloatBarrier
\bibliographystyle{JHEP.bst}
\bibliography{main}

\end{document}